\documentclass[a4paper,twocolumn,10pt,noarxiv,accepted=2021-09-27]{quantumarticle}
\pdfoutput=1
\usepackage[utf8]{inputenc}
\usepackage[table]{xcolor}
\usepackage{graphicx}
\usepackage{amsmath}
\usepackage{amssymb}
\usepackage{svg}
\usepackage{physics}
\usepackage{gensymb}
\usepackage{verbatim}
\usepackage{setspace}
\usepackage{graphicx}
\usepackage{array}
\usepackage{makecell}
\usepackage{float}
\usepackage{hyperref}
\usepackage[numbers,sort&compress]{natbib}

\makeatletter
\renewcommand{\fnum@figure}{\textbf{Fig.\,\thefigure}}

\makeatother

\begin{document}

\title{Multidimensional cluster states using a single spin-photon interface coupled strongly to an intrinsic nuclear register}

\author{Cathryn P. Michaels\textsuperscript{1,*}}
\author{Jes\'{u}s Arjona Mart\'{i}nez\textsuperscript{1,*}}
\author{Romain Debroux\textsuperscript{1,*}}
\author{Ryan A. Parker\textsuperscript{1}}
\author{Alexander M. Stramma\textsuperscript{1}}
\author{Luca I. Huber\textsuperscript{1}}
\author{Carola M. Purser\textsuperscript{1}}
\author{Mete Atat\"ure\textsuperscript{1,$\dagger$}}
\author{Dorian A. Gangloff\textsuperscript{1,$\dagger$}}

\affiliation{\textsuperscript{1} Cavendish Laboratory, University of Cambridge, JJ Thomson Avenue, Cambridge, CB3 0HE, UK
\\
\textsuperscript{*}\,These authors contributed equally to this work.
\\
\textsuperscript{$\dagger$}\,Correspondence should be addressed to: ma424@cam.ac.uk, dag50@cam.ac.uk.
}

\date{\today}

\begin{abstract}
    Photonic cluster states are a powerful resource for measurement-based quantum computing and loss-tolerant quantum communication. Proposals to generate multi-dimensional lattice cluster states have identified coupled spin-photon interfaces, spin-ancilla systems, and optical feedback mechanisms as potential schemes. Following these, we propose the generation of multi-dimensional lattice cluster states using a single, efficient spin-photon interface coupled strongly to a nuclear register. Our scheme makes use of the contact hyperfine interaction to enable universal quantum gates between the interface spin and a local nuclear register and funnels the resulting entanglement to photons via the spin-photon interface. Among several quantum emitters, we identify the silicon-29 vacancy centre in diamond, coupled to a nanophotonic structure, as possessing the right combination of optical quality and spin coherence for this scheme. We show numerically that using this system a $2$$\times$$5$-sized cluster state with a lower-bound fidelity of $0.5$ and repetition rate of $65$\,kHz is achievable under currently realised experimental performances and with feasible technical overhead. Realistic gate improvements put $100$-photon cluster states within experimental reach.
\end{abstract}

\maketitle

\section{Introduction}
Photons offer a robust way to distribute information and entanglement to implement building elements of quantum communication and computation~\cite{Aspect1981,Ekert1991,Bouwmeester1997,Raussendorf2001a,Raussendorf2003,Briegel2009a,Kimble2008,Ladd2010}. Photon loss is the primary culprit for rate and fidelity reduction in photonic systems \cite{Gisin2002,Zhong2020,Xu2020}. A remedy to photon loss is to utilise multiple photons in a way that makes the joint state resilient to photon loss. Cluster states are a promising solution for this approach as they possess two key properties: (1) persistency, meaning entanglement remains even if individual photons are lost or measured, and (2) maximal connectedness, meaning any two qubits can be brought into a Bell state via local measurements~\cite{Briegel2001, Varnava2008, Zwerger2016}. In particular, multidimensional cluster states enable universal measurement-based quantum computation~\cite{Raussendorf2001a, Raussendorf2003, Briegel2009a} and measurement-based quantum repeaters for long range quantum communication~\cite{Zwerger2016, Azuma2015All-photonicRepeaters}. Photonic cluster states have been realised via both probabilistic entanglement schemes, using spontaneous parametric downconversion~\cite{Grice2011ArbitrarilyElements, Kilmer2019BoostingDetectors, Ewert20143/4Ancillae,Browne2005Resource-EfficientComputation,Zhao2004ExperimentalTeleportation, Gao2010ExperimentalState, Wang2016ExperimentalEntanglement} and delay lines \cite{Istrati2019}, as well as via deterministic entanglement schemes using squeezed light and non-gaussian measurements \cite{Asavanant2019}, and using quantum-dot (QD) spin to single-photon interfaces \cite{Lindner2009a,Schwartz2016}. 

A deterministic approach was proposed initially for microwave photons coupled to Rydberg atoms \cite{Gontagonta2009}. Simultaneously, an approach for optical photons coupled to a solid-state emitter was proposed~\cite{Lindner2009a} as a more feasible scheme to generate one-dimensional photonic cluster states using a single spin-photon interface and is of particular interest. In this approach, an optically addressable spin emits photons sequentially, which are entangled to one another via gates performed on the electron spin between each photon-emission event. In 2010, Economou~\textit{et~al.} extended this Lindner-Rudolph scheme by using multiple coupled spin-photon interfaces to create a multi-dimensional cluster state \cite{Economou2010b}, necessary for quantum information processing~\cite{Mantri2017}. This proposal has since been extended to leverage more recent experimental advances on interactions between QDs~\cite{Gimeno-Segovia2018}, and to produce arbitrary graph states from linear arrays of spin-photon interfaces~\cite{Russo2019}.

\begin{figure*}[t]
    \centering
    \includegraphics[width=0.9\textwidth]{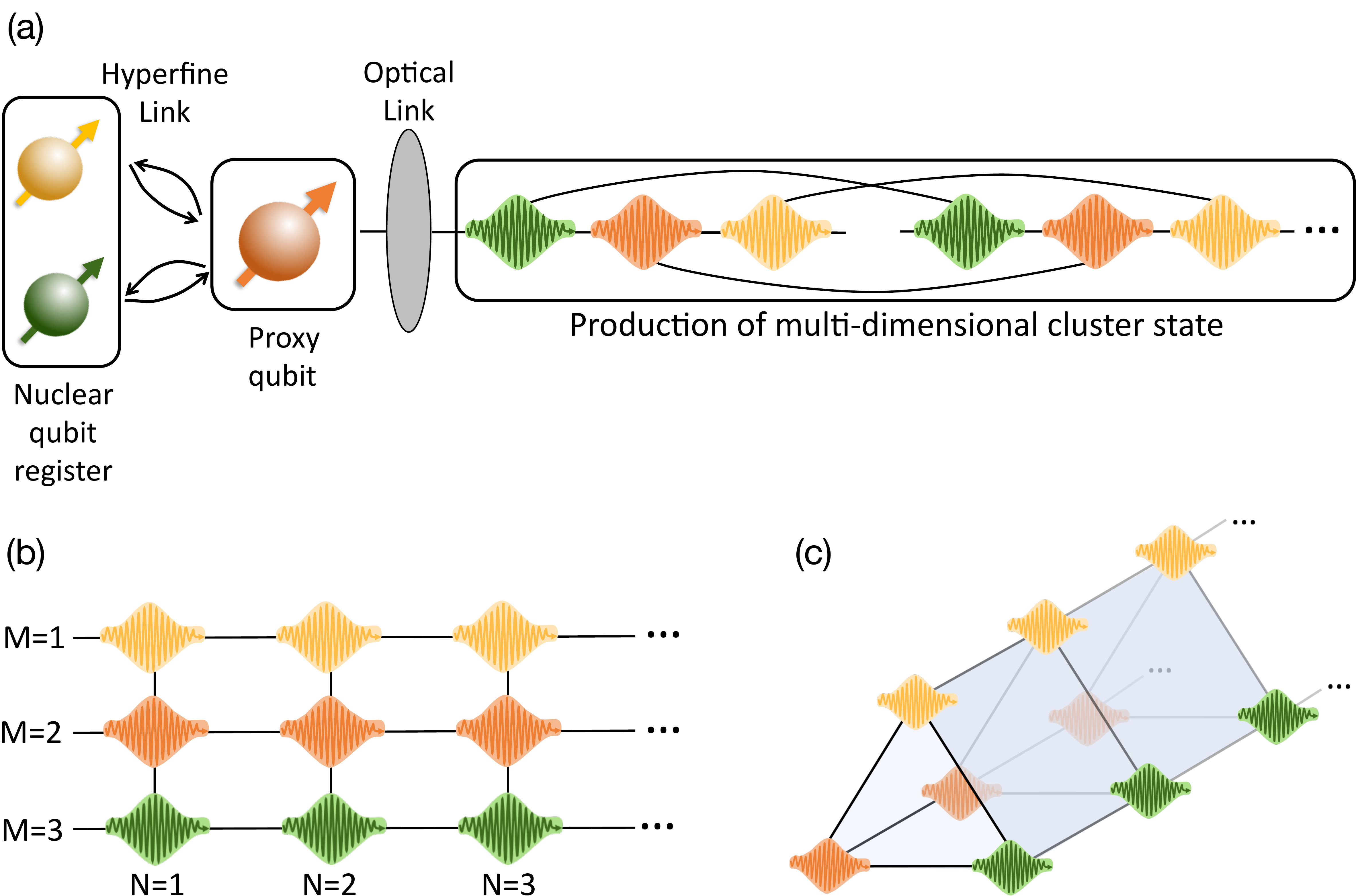}
    \caption{Multi-dimensional entangled state production. (a) Linear production of entangled photons via an optical link with a single proxy qubit. The hyperfine link between nuclear register and proxy qubit provides additional entanglement dimensions amongst the linear string of photons. (b) 2-dimensional cluster state equivalent to one produced via the method shown in (a). (c) 3-dimensional \textit{Toblerone} state which could be produced by extending the scheme in (a), entangling the nuclear qubits to each other.}
    \label{fig1}
\end{figure*}

The complex photonic states necessary for quantum communication, named ``repeater graph states"~\cite{Azuma2015All-photonicRepeaters}, can also be produced using two QDs~\cite{Russo2018PhotonicCommunications} as well as a single optically active spin coupled to an ancilla qubit, such as a proximal nuclear spin~\cite{Russo2018PhotonicCommunications,Buterakos2017DeterministicEmitters}.
The latter approach is particularly attractive because of the reduced technical overhead with respect to engineering interactions between multiple quantum emitters. The ideal candidate is therefore an optically active spin that combines excellent optical properties, long spin coherence, and an intrinsic (deterministic) nuclear register strongly coupled to the spin for fast exchange of information.

In this proposal, we provide a scheme to generate a multidimensional cluster state from a single spin-photon interface coupled strongly to an intrinsic nuclear register via the hyperfine interaction. This scheme can leverage a single high-spin nucleus acting as a multi-qubit nuclear register. We identify solid-state systems which could be used to realise this scheme via an intrinsic nuclear spin, with group IV colour centres in diamond constituting a particularly promising platform. We analyse as an example case the better-studied of these, the silicon vacancy (SiV) colour centre in diamond coupled to a spin-$\frac{1}{2}$ nucleus. We calculate the state fidelity as a function of cluster state size for a range of material parameters and find that, when limited by gate fidelity values already verified experimentally, our scheme can generate at a rate of 65\,kHz a $10$-photon cluster state of dimensions $2$$\times$$5$ with fidelity $F > 0.5$ – exceeding the current state-of-the-art for cluster states based on single photons. With realistic gate improvements, this can be extended to generating a $100$-photon, $2$$\times$$50$ cluster state at 0.6\,mHz  with $F > 0.9$.

\begin{figure*}
    \includegraphics[width=\textwidth]{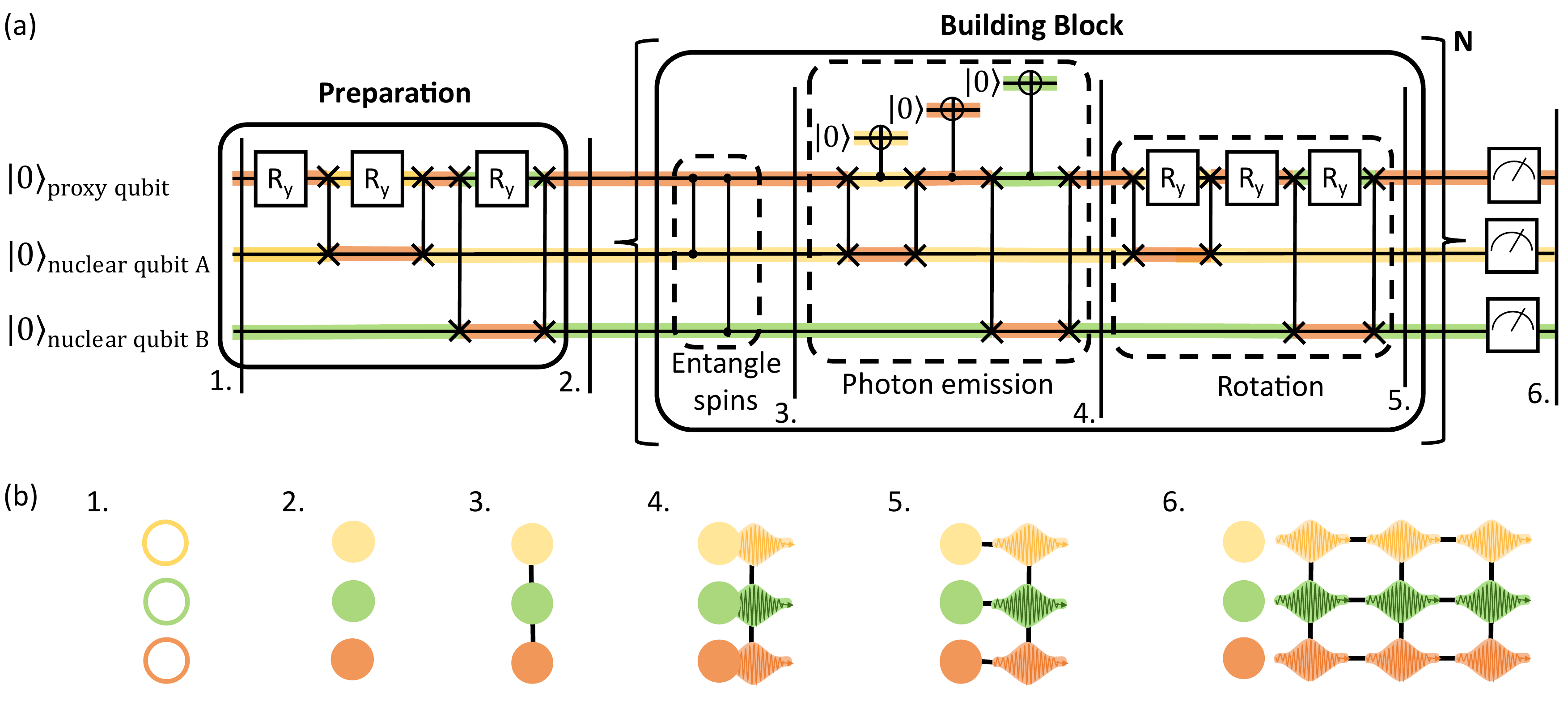}
    \hfill
    \caption{(a) Quantum circuit and (b) diagram outlining the generation of a $3$$\times$$N$ sized cluster state. The initialisation block is followed by the building block, repeated $N$ times. The coloured lines represent the position of each information qubit along the circuit rails, swapping positions on a SWAP gate and showing which information qubit each gate is acting on.}
    \label{fig2}
\end{figure*}

\section{Cluster State Generation}
Figure\,\ref{fig1}a illustrates the basic elements of our protocol, allowing for tailored connectedness among photons. A hyperfine link repeatedly creates entanglement between the multi-qubit nuclear register and the proxy qubit. One-by-one, the optical link generates photon qubits which are entangled with the proxy qubit, using any of a polarisation, a frequency, or a time-bin scheme. Figure~\ref{fig1}b shows that by unfolding the linear chain of photons in Fig.~\ref{fig1}a, we reveal the two-dimensional nature of the entanglement connections within the cluster state. By extension, any $M \times N$ two-dimensional cluster state can be produced using a proxy qubit and $M-1$ qubits in the nuclear register, each linked to $N$ photons. The outer qubit rails of Fig.~1b can be linked further using the hyperfine interaction to create a three-dimensional structure, such as the \textit{Toblerone} state of Fig.~\ref{fig1}c. 

Figure~\ref{fig2}a shows a quantum circuit that generates a two-dimensional, $3\times N$ cluster state in our approach, broken down into a preparation block, performed once, and a building block, repeated $N$ times. All single-qubit gates are performed on the proxy qubit, with hyperfine-enabled SWAP gates used to transfer the nuclear qubit information to and from the proxy qubit. Figure~\ref{fig2}b shows a schematic diagram that illustrates the following detailed stages of the protocol, using the computational basis $\{\ket{0},\ket{1}\}$ as qubit states:

\begin{enumerate}
    \item \textit{Initialisation}: the proxy and nuclear qubits start in the $\ket{0}$ state.
    
    \item \textit{Preparation}: the proxy and nuclear qubits are each rotated, using a $\pi/2$ y-axis rotation ($\rm{R_y}$) gate, into an identical superposition state of $\ket{0}$ and $\ket{1}$.
    
    \item \textit{Qubit entanglement}: hyperfine-enabled controlled-phase (CZ) gates are applied sequentially to entangle each nuclear qubit to the proxy qubit.
    
    \item \textit{Photon emission}: cycling through each qubit in the register (via SWAP gates), the proxy qubit generates a photon qubit whose state is entangled with that of the proxy qubit; this is equivalent to a controlled-not (CNOT) gate. 
    
    \item \textit{Rotation}: $\rm{R_y}$ gates extend the size of the cluster state along $N$ and prepare the nuclear and proxy qubits for the next iteration of the building block. \newline
    
    Return to \textit{Qubit entanglement} (stage 3) $N-1$ times.
    \newline
    
    \item \textit{Completion}: measuring the proxy and nuclear qubits in the \textit{z} basis results in a purely photonic cluster state of $M\times N$ photons.
\end{enumerate}
\noindent
The resulting time-ordered $M$$\times$$N$ photonic cluster state is then ready to serve as the input state for a measurement-based quantum algorithm. We show the detailed quantum-state tracking and its equivalence to a cluster state in Appendix\,\ref{ap:state}. We note that the presented circuit is constructed for maximal pedagogical value, and there exists optimised circuits with a reduced number of rate-limiting SWAP gates, resulting in the same cluster state.

\section{Suitable systems for implementation}

To realise this protocol with high fidelity and high production rate the system of choice must be a bright and coherent photon source with a coherent electronic ground-state spin, which in turn couples strongly to a nuclear register. The brightness and coherence of the photon source can be quantified according to its excited state lifetime ($\tau$), its internal quantum efficiency ($\eta_\text{QE}$), overall photon collection efficiency ($\eta_\text{CE}$) and Debye-Waller factor ($\eta_\text{DWF}$). It must also possess cyclic optical transitions that can be used to generate entanglement between the ground state spin and the photons in a chosen degree of freedom, such as frequency or polarisation. The coherence time of both the ground-state spin ($T_2$) and the nuclear register must be long compared to the duration of the whole protocol for the desired value of $N$. This duration is dominated by the strength of the hyperfine interaction ($A$), which limits the speed of SWAP and CZ gates between the proxy qubit and the nuclear register. This register can be realised through proximal nuclear spins \cite{waldherr2014quantum}, collective excitations of a nuclear spin ensemble \cite{gangloff2019nuclearensemble}, or a single intrinsic nuclear spin ($I \ge \frac{1}{2}$) \cite{metsch2019nuclearsiv}. Given the importance of efficient generation and collection of photons, optically active spins in solids~\cite{Atature2018} offer the advantage of being more readily coupled to nanostructures, enabling higher collection efficiencies \cite{janitz2020cavity,OBrien2009}. Table\,1 presents a selection of such candidate systems for a comparison across all relevant and available metrics.

\newcommand{\colorA}{EEF9E2}
\newcommand{\colorB}{FFF5DB}
\newcommand{\colorC}{F9EBE1}

\newcolumntype{N}{@{}m{0pt}@{}}

\renewcommand\theadalign{tc}
\renewcommand\theadfont{\normalsize}

\begin{table*}
\renewcommand{\arraystretch}{1.0}
\setlength{\tabcolsep}{1pt}
\begin{tabular}{
>{\columncolor[HTML]{\colorA}\centering\arraybackslash}p{14mm}
>{\columncolor[HTML]{\colorB}\centering\arraybackslash}p{16mm}
>{\columncolor[HTML]{\colorB}\centering\arraybackslash}p{16mm}
>{\columncolor[HTML]{\colorB}\centering\arraybackslash}p{16mm}
>{\columncolor[HTML]{\colorB}\centering\arraybackslash}p{16mm}
>{\columncolor[HTML]{\colorB}\centering\arraybackslash}p{16mm}
>{\columncolor[HTML]{\colorC}\centering\arraybackslash}p{16mm}
>{\columncolor[HTML]{\colorC}\centering\arraybackslash}p{16mm}
>{\columncolor[HTML]{\colorC}\centering\arraybackslash}p{38mm}N}

 & $\tau$ (ns) & $C$ & $\eta_\text{QE}$ & $\eta_\text{CE}$ & $\eta_\text{DWF}$ & $T_2$ ($\mu$s) & $A$ (MHz)  & $I$ &\\[2pt] \hline
 QD & 0.6-0.8 \cite{Paillard2000}  & \thead{150 \\ \cite{najer2019gated}}  & \thead{$\sim$ 1 \\ \cite{Senellart2017}} & \thead{79\% \\ \cite{Senellart2017}} &  88-95\% \cite{peter2004phonon, matthiesen2013phase, konthasinghe2012coherent}   & \thead{3 \\ \cite{Bechtold2015,Stockill2016}} & $1$-$5^{\dagger\dagger}$  \cite{Hogele2012a,gangloff2019nuclearensemble} & $I=3/2$  ($^{69}$Ga,~$^{71}$Ga,~$^{75}$As), $I=9/2$ ($^{173}$In,$^{175}$In) &\\[5pt] 
 SiC (VV$^0$) & \thead{15.8-18.7 \\ \cite{christle2017isolated}}  & - & - & \thead{$\sim$40\% \\ \cite{calusine2014silicon}} &  \thead{7\% \\  \cite{christle2017isolated}}  &  $1.5 \times 10^4$ \cite{Bourassa2020}  & \thead{13.2 \\ \cite{Bourassa2020}}  & $I=1/2$ ($^{29}$Si)&\\[5pt] 
 SiC (V$^{4+}$) & \thead{11 \\ \cite{Spindlberger2019}} &  - & - & \thead{$\sim$40\% \\ \cite{calusine2014silicon}} &  \thead{50\% \\ \cite{wolfowicz2020vanadium}}   & - & \thead{232 \\ \cite{wolfowicz2020vanadium}} & $I=7/2$ ($^{51}$V) &\\[5pt]  
 NV & \thead{13 \\ \cite{manson2006nvstructure}} & $0.03$ \cite{riedel2017deterministic,janitz2020cavity} & \thead{0.5-1 \\ \cite{Berthel2015}} & \thead{37\% \\ \cite{Patel2016}} &   \thead{4\% \\ \cite{aharonovich2011diamond}}    & $6.8 \times 10^5$ \cite{humphreys2017deterministicentanglement} & \thead{2.18 \\ \cite{Pfaff2013}} & \thead{$I = 1$ ($^{14}$N), \\ $I = 1/2$ ($^{15}$N)} & \\[5pt]
 SiV & \thead{1.6 \\ \cite{becker2018siv}} & \thead{105 \\ \cite{bhaskar2020memoryenhanced}}  & \thead{0.1 \\ \cite{sukachev2017siv}} & \thead{85\% \\ \cite{bhaskar2020memoryenhanced}} & 70-80\%  \cite{Neu2011FluorescenceIridium,Neu2011} & $1.0 \times 10^4$ \cite{sukachev2017siv} & \thead{70 \\ \cite{Pingault2017CoherentDiamond,Edmonds2008ElectronDiamond}} & $I=1/2$ ($^{29}$Si)&\\[5pt] 
 GeV & \thead{1.4–5.5 \\ \cite{Iwasaki2015a}}&  \thead{0.1 \\  \cite{bhaskar2017quantum}} & \thead{0.4 \\ \cite{bhaskar2017quantum}} & 85\%$^\dagger$ &  60\% \cite{Palyanov2015,bhaskar2017quantum}   &- & - & $I=9/2$ ($^{73}$Ge)&\\[5pt] 
 SnV & \thead{4.5 \\ \cite{trusheim2020transformlimitted}} & \thead{9$^*$ \\ \cite{rugar2021quantum}} & \thead{0.8 \\ \cite{Iwasaki2017}} & 85\%$^\dagger$ &  \thead{60\% \\ \cite{gorlitz2020spectroscopic}}  & \thead{$300$ \\ \cite{Debroux2021}} & \thead{$42.6$ \\ \cite{Debroux2021}}  & $I=1/2$ ($^{115}$Sn,~$^{117}$Sn,~$^{119}$ Sn)&  \\[5pt]
 
 \end{tabular}
 \caption{Parameters: $\tau$ excited state lifetime, $C$ cooperativity (best to date), $\eta_\text{CE}$ collection efficiency (best to date), $T_2$ coherence time (with dynamical decoupling), $A$ hyperfine constant, $I$ intrinsic nuclear spin. $^\dagger$ expected value. $^{\dagger\dagger}$ collective non-collinear interaction. $^*$ estimated, corrected for zero-phonon line emission. }
 \label{tab: proposed systems}
\end{table*}

Self-assembled InGaAs QDs remain the best performing single-photon sources, featuring near-unity quantum efficiency and record-high photon collection via cavity coupling \cite{Senellart2017,najer2019gated,Tomm2021}, as well as fast and high-fidelity control of the electron spin \cite{Kim2011UltrafastSpins, Ding2019CoherentInterface,bodey2019spinlocking}. They remain the only spin-photon interface that has been used to generate photonic cluster states \cite{Schwartz2016}. They also feature a multi-mode nuclear spin ensemble that could serve as a large nuclear spin register for this scheme \cite{Denning2019}. However, reaching beyond proof-of-concept demonstrations faces the challenge of modest proxy spin coherence times, which can be up to a few microseconds \cite{Bechtold2015,Stockill2016}. 

Colour centres are another realisation of single-photon sources in the solid-state. Those in silicon carbide have access to proximal nuclear spins and in the VV$^0$ colour centre, CNOT gates have been realised, as well as nuclear-electron spin SWAP gate fidelities of greater than 93\% \cite{DeLasCasas2017,Bourassa2020}. The main challenge lies in improving collection efficiencies \cite{calusine2014silicon}, and this will depend on their spin and photon quality once they are coupled to photonic nanostructures. We also note centres in other systems, including defects in hexagonal boron nitride \cite{Tran2016} and rare-earth ions \cite{zhong2017}, which can also be suitable for photonic cluster state generation, after we gain sufficient insight into their optical and spin properties in tandem. 

Diamond is a particularly promising host for colour centres, possessing the largest optical bandgap of any material \cite{Aharonovich2011, Aharonovich2014a, Aharonovich2016}. The most understood colour centre in diamond is the nitrogen vacancy (NV) centre displaying outstanding spin coherence even at room temperature. The intrinsic host nuclear spin, $^{14}\text{N}$, features a 2.18~MHz hyperfine coupling strength \cite{Pfaff2013} and can be addressed both through frequency-selective microwave pulses \cite{Pfaff2013} and via the coupling of spin sub-levels \cite{fuchs2011quantummemory,holzgrafe2019correctedreadout}. Its state-selective optical transitions have also been used to demonstrate spin-photon entanglement \cite{Togan2010}. The challenge lies in improving the fraction of photons emitted and collected via cavity coupling, but progress in this area has so far remained modest due to the degradation of the spin and optical qualities of the NV centre in the vicinity of nanostructured surfaces \cite{janitz2020cavity}.

The more recently studied group IV colour centres in diamond possess an inversion-symmetric molecular structure that makes them less sensitive to electric field fluctuations, ultimately resulting in improved optical properties and compatibility with diamond nanostructures \cite{Bradac2019a, Iwasaki2015a, trusheim2020transformlimitted, Trusheim2019, Wan2019}, such as waveguides and nanocavities, achieving full system detection efficiencies ($\eta_{CE}$) of $85\%$ \cite{bhaskar2020memoryenhanced} and cavity-coupling efficiencies ($\beta$-factor) of up to $95\%$ \cite{rugar2021quantum,kuruma2021,fuchs2021}. Group IV colour centres have also been coupled to nanocavities with cooperativity $C = 105 \pm 11$, resulting in Purcell enhancement of the radiative decay paths and an improvement of $\eta_\text{QE}$ \cite{bhaskar2020memoryenhanced}. The SiV group IV colour centre has $\eta_\text{DWF}=0.7$-$0.8$\,\cite{Neu2011FluorescenceIridium,Neu2011}, whilst the tin and germanium vacancies have been measured at $\eta_\text{DWF} = 0.6$\,\cite{Palyanov2015,gorlitz2020spectroscopic}.  Of the group IV colour centres, the most widely studied has been SiV \cite{Hepp2014, rogers2014siv, sukachev2017siv, Pingault2017CoherentDiamond,  becker2018siv, meesala2018strainengineering, metsch2019nuclearsiv, trusheim2020transformlimitted, Sohn2018a}, for which a dynamically decoupled coherence time $T_2$ has been shown to exceed $10$\,ms at 100\,mK\,\cite{sukachev2017siv}. $^{29}\text{Si}$ can provide an intrinsic  nuclear spin, and has been measured to have an isotropic hyperfine interaction with $A_\perp \approx A_\parallel = 70$ MHz \cite{Pingault2017CoherentDiamond}, in agreement with theoretical predictions \cite{Gali2013}. Whilst $^{29}\text{Si}$ is a $\frac{1}{2}$ nuclear spin, $^{73}\text{Ge}$ is a $\frac{9}{2}$ nuclear spin, allowing for a multi-mode (or multi-qubit) nuclear register. The remaining group IV colour centres are expected to follow similar trends in their optical and spin properties, making them a natural platform on which to perform our multidimensional cluster state generation scheme. 

\section{Cluster-state generation with SiV centres}

\begin{figure}
    \centering
    \includegraphics[width=\columnwidth]{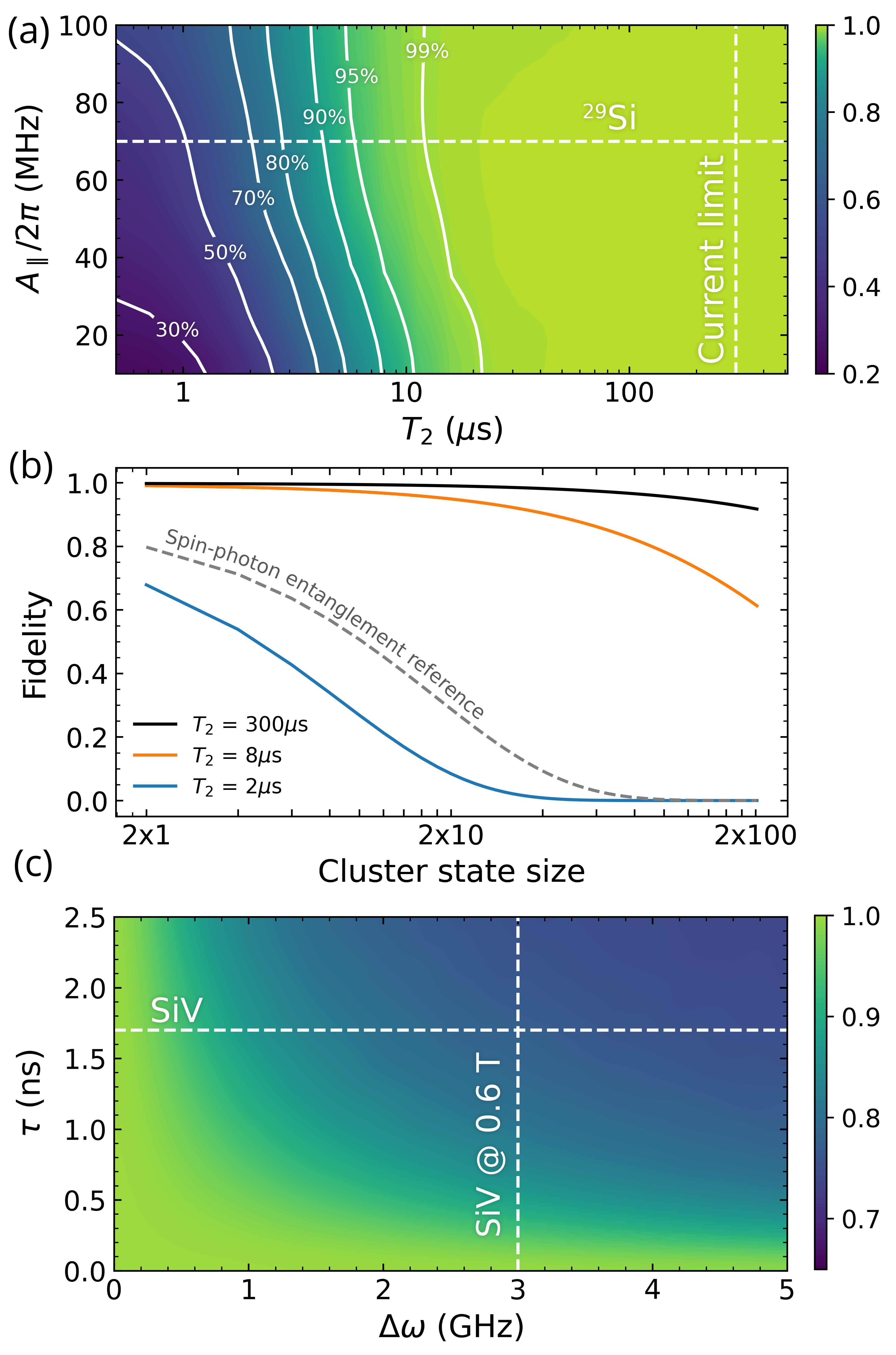}
    \caption{
    (a) Calculated fidelity of a $2$$\times$$2$ cluster state as a function of the electronic coherence time $T_2$ (Hahn echo) and the parallel component of the hyperfine interaction $A_\parallel$. At each point,  the magnetic field is selected to minimise the time-length of the sequence. The effects of finite $T_2$ are introduced via an Ornstein–Uhlenbeck noise bath (Appendix\,\ref{ap:simulation}). The dashed lines indicate parameters relevant to $^{29}\text{Si}$: $A_\parallel = 70$\,MHz and $T_2 = 300$\,$\mu$s. \cite{Pingault2017CoherentDiamond, sukachev2017siv} (b) Extrapolated decoherence-limited fidelity of a cluster state as a function of the cluster state length for $T_2 = 2$\,$\mu$s (blue), $T_2 = 8$\,$\mu$s (orange), and for $T_2 = 300$\,$\mu$s (black). The dashed grey line shows the fidelity limit imposed by a $94\%$-fidelity spin-photon entanglement gate. (c) Fidelity of the photon emission as a result of the spin dephasing in the excited state, shown as a function of the lifetime of the excited state ($\tau$) and of the difference in electronic precession frequency between the ground and optically excited states ($\Delta \omega$). The dashed lines indicate parameters relevant to our simulation of SiV: $\tau = 1.7$\,ns and $ \Delta \omega = 3$\,GHz.}
    \label{fig:GRAPE}
\end{figure}

To test the feasibility of our proposed scheme we estimate a lower-bound for the fidelity and rate with which our scheme could be used to produce a 2D cluster state using an SiV centre with an intrinsic spin-$\frac{1}{2}$ $^{29}\text{Si}$ nucleus. To do so, the conceptual circuit shown in Fig.\,2 must be translated into a physical pulse sequence tailored for the SiV transitions, for which: (i) control of the electron qubit has been achieved with both microwave \cite{Pingault2017} and all-optical \cite{becker2018siv} pulses in resonance with the qubit splitting, (ii) two-qubit gates can be performed solely through control of the electron by exploiting its hyperfine interaction with the ancillary qubit \cite{metsch2019nuclearsiv}, and (iii) the existence of cyclic transitions between the ground and excited states \cite{sukachev2017siv} can allow spin-dependent photon emission.  

\emph{Electron-nuclear gates --} The link between the electronic and the $^{29}\text{Si}$ nuclear spins in SiV is an isotropic hyperfine interaction and thus of the form $A_\parallel \vec{\sigma} \cdot \vec{\sigma}$, where $A_\parallel = A_\perp = 70$\,MHz \cite{Pingault2017CoherentDiamond} is the hyperfine constant and $\vec{\sigma}$ are the Pauli matrices. Under an external magnetic field the splitting between the electronic states can be made larger than this hyperfine interaction owing to the electronic gyromagnetic ratio ($12$-$25$\,GHz/T \cite{rogers2014siv,Pingault2017CoherentDiamond,sukachev2017siv}), and thus $A_\parallel \sigma_z \otimes \sigma_z$ can be treated as the dominant secular term, where $z$ is the quantisation axis of the electron. In the regime where the spin-orbit coupling rate $\lambda_\text{SO}$~=~50\,GHz \cite{Hepp2014} dominates over the Zeeman splitting of the electron, the quantisation axis ($z$) remains predominantly along the symmetry axis of the SiV defect (see also Appendix\,\ref{ap:g4hamil}).

The $A_\parallel \sigma_z \otimes \sigma_z$ term acts as a conditional effective magnetic field on the nucleus, dependent on the state of the electron, which can be exploited to realise two-qubit gates between the electron and nuclear spins. Dynamical decoupling gates built on this conditionality have been used to control $^{13}$C spins in proximity to both the NV and the SiV centres \cite{taminiau2014universalcontrol, bhaskar2020memoryenhanced}. For $^{13}$C, these exploit the dipolar coupling of the two spins, which takes the form $\sigma_z \otimes \sigma_x$. To control the intrinsic $^{29}\text{Si}$ nuclear spin an analogous term can be achieved with an off-axis magnetic field, which dominates as the quantisation axis of the nucleus, but is not strong enough to overcome the electron's spin-orbit coupling. This results in axes of precession for the nucleus that are conditional on the electron state \cite{taminiau2014universalcontrol}. Dynamically decoupled protocols of the form $(\tau_f - \pi - 2\tau_f - \pi - \tau_f)^N$, where $\tau_f$ corresponds to a free precession time and $\pi$ to an electronic $R_x(\pi)$ gate, can then be used. By matching the interpulse spacing $2\tau_f$ to the nucleus precession dynamics, conditional and unconditional rotations of the nuclear spin can be realised. By combining these units with gates on the electron, the required SWAP and CZ two-qubit operations, as well as single nuclear qubit operations can be realised. 

\emph{Spin-operations fidelity --} High-fidelity electron gates ($>99\%$) and spin-photon entanglement ($>94\%$) have both been achieved in SiV \cite{bhaskar2020memoryenhanced}, and neither are limited by fundamental properties of the light-matter interface. As such, we characterise the errors induced by electron-nuclear gates by first simulating the cluster state as a whole considering instantaneous, ideal single-qubit gates on the electron and ideal single photon emission. Under this assumption, we simulate the Fig.~2a circuit output for a $2$\,$\times$\,$2$ cluster state,  $\rho$, and calculate the fidelity $F_{2\times 2} = \sqrt{\bra{\psi_{2\times 2}} \rho \ket{\psi_{2\times 2}}}$ of a cluster state $\ket{\psi_{2\times 2}}$; our simulation procedure is explained in detail in Appendix \ref{ap:simulation}. Due to the small value of the nuclear Zeeman splitting compared to typical hyperfine constants, a relatively large magnetic field is preferable to obtain fast gates. Here we select $B_z$=0.6\,T and $B_x$=0.6\,T to maintain achievable microwave control frequencies with gate durations of 1.6\,$\mu$s for SWAP gates and 1.1\,$\mu$s for CZ gates. We identify the electron spin coherence as the dominant dephasing term in the electron-nuclear interaction, as characterised by the Hahn-echo $T_2$ whose value is limited by nuclear spin noise and sample temperature \cite{sukachev2017siv}. Figure\,3a shows the calculated fidelity $F_{2\times 2}$ as a function of the electronic $T_2$ (Hahn echo) and the hyperfine constant $A_\parallel$, which allows straightforward comparison with similar group-IV systems.  Below $T_2 \sim 8$\,$\mu$s, we observe an increase of fidelity with hyperfine constant, owing to faster gates. Above $T_2 \sim 8$\,$\mu$s, we obtain a state fidelity of 99.9\%, a constant value which is limited only by our gate optimisation procedure (Appendix\,\ref{ap:simulation}). While more advanced pulse schemes, such as PulsePol for SWAP gates \cite{Schwartz2018}, could result in a shorter sequence length and improved fidelities, our proposed dynamically decoupled gates provide a baseline of the expected decoherence-induced infidelities. 

\emph{Extended cluster state fidelity --} To estimate the fidelity of longer cluster states, we take the loss of fidelity within each part of the scheme, preparation and building block, finding fidelities of 99.9\% and 99.8\% respectively. We extrapolate this loss exponentially to the target cluster state length, assuming the electron spin initialisation and readout in stages 1 and 6 in Fig.\,\ref{fig2} have near-unity fidelity \cite{bhaskar2020memoryenhanced}. Figure 3b shows the cluster state fidelity $F$ as a function of its length for several values of $T_2$ from $2$\,$\mu$s to $300$\,$\mu$s. We also show the limit set by, to date, the best achieved spin-photon entanglement fidelity of 94\% (grey dashed line). Under such experimental conditions, our scheme can generate a $2$\,$\times$\,$5$ cluster state with $F>0.5$. In stark contrast, spin-photon entanglement operations of unity fidelity combined with $T_2 \sim 300$\,$\mu$s allows for the same cluster state to be produced with $F>0.99$, and for $2$\,$\times$\,$50$ cluster states with $F>0.90$.

\emph{Dephasing errors in photon emission --} Errors in the photon emission step depend on the approach used to transfer the spin qubit onto the photon. A natural spin-photon entanglement scheme to consider is to excite two cyclic optical transitions from the ground state to the excited state of SiV simultaneously \cite{sukachev2017siv}, as is in the original Lindner et al.~proposal \cite{Lindner2009a}. However, at finite magnetic field $|B|$ required by our scheme, the non-degenerate frequencies of these two optical transitions introduce a spin phase uncertainty of $\phi = \Delta g_\text{e} \mu_\text{B} |B| \tau$, where $\tau$ is the finite lifetime of the excited state, and $\Delta g_\text{e}$ the electronic g-factor difference between the ground and excited states. To quantify the corresponding error on the spin-photon entangled state, we consider an initial electronic spin state $(\ket{0} + \ket{1})/\sqrt{2}$, which after optical excitation and subsequent photon emission transforms to $\ket{\Psi(\phi)} = (\ket{0}\ket{\gamma_0} + e^{\rm{i} \phi}\ket{1}\ket{\gamma_1})/\sqrt{2}$, where $\ket{\gamma_{0,1}}$ are the states of the emitted photons. Figure 3c shows the  fidelity $F = \text{max}_{\ket{\alpha}} \sqrt{\bra{\alpha} \rho \ket{\alpha}}$, where $\ket{\alpha}$ is a pure state and $\rho$ is the density matrix obtained by averaging over the distribution of states $\ket{\Psi(\phi)}$. As an example point (dashed lines), we see that for the combination of the SiV natural excited state lifetime and a magnetic field that is specific to our cluster state calculations, the spin-photon entanglement fidelity would be limited to $\sim 80\%$.

Such an error process is general to schemes that encode photon qubits as colour or polarisation. In the former case, the requirement to resolve the two transitions in colour imposes $\phi = \Delta g_\text{e} \mu_\text{B} \lvert B \rvert \tau \gtrsim 2\pi$, which necessarily introduces a large unknown dephasing. This effect has been observed as an oscillation in the qubit readout as a function of the photon's emission time \cite{Togan2010, degreve2012, gao2012,Schaibley2013}. Polarisation encoding is thus by far preferable as it imposes no restriction on the excited state lifetime, which can be made short enough (e.g. by Purcell enhancement) such that $\phi = \Delta g_\text{e} \mu_\text{B} \lvert B \rvert \tau \ll 2\pi$, and is the commonly employed basis in proposals \cite{Russo2018PhotonicCommunications} and experimental realisations in QDs \cite{Schwartz2016} and NVs \cite{Togan2010}. In the case of SiV, such a polarisation encoding is possible in unstrained centres, for which optical emission between states of orthogonal orbitals corresponding to the spin-flipping peaks is circularly polarized ($\sigma^+, \sigma^-$), and is the only emission collected along the SiV $\langle 111 \rangle$~symmetry axis \cite{Hepp2014}. To achieve high collection efficiency, the spin-flipping optical transitions must be nearly cyclic, which could be achieved using a cavity oriented along the $\langle 111 \rangle$~diamond orientation. 

As a third spin-photon entanglement scheme to consider, the photon qubit can be encoded in time bins. This is achieved by exciting a single, cyclic spin conserving transition, followed by applying a $\pi$ pulse to invert the qubit, exciting the same transition again, and finally applying a $\pi$ pulse to restore the spin qubit. In this encoding, neither polarisation selective optical rules nor energy-degenerate transitions are required. Further converting the time bins into polarisation degrees of freedom bypasses the issue of excited state dephasing presented above and has been demonstrated with an NV center \cite{vasconcelos}.

\emph{Cluster state generation rate --} The rate $R$ at which a cluster state containing $M \times N$ photons can be produced and used is exponentially dependent on the generation, collection, and detection ($\eta_\text{DE}$) efficiencies: $R~\propto~\left(\eta_\text{QE}\cdot \eta_\text{DWF}\cdot \eta_\text{CE}\cdot \eta_\text{DE}\right)^{M \times N}$. Full system efficiencies $\eta_\text{QE}\cdot \eta_\text{DWF}\cdot \eta_\text{CE}\cdot \eta_\text{DE} =  85\%$ have been shown for diamond cavities with cooperativity of $C = 105 \pm 11$ \cite{bhaskar2020memoryenhanced}, enabled by high-efficiency diamond-fibre coupling ($>$90$\%$) and routing to superconducting nanowire single photon detectors. In this case, a $2$$\times$$5$ cluster state (scheme length of $3$\,$\mu$s) can be produced at 65\,kHz and a $2$$\times$$50$ cluster state (scheme length of $30$\,$\mu$s) can be produced at 0.6\,mHz.

~\newline
\section{Conclusions}

We have presented a protocol to generate multidimensional photonic cluster states using a single spin-photon interface by making use of the strong hyperfine link between an optically active electronic spin and an intrinsic nuclear spin register. We explored the feasibility of our scheme via the diamond $^{29}$Si-vacancy centre coupled to a nanophotonic structure, and used current experimental performance to show that a $2$$\times$$5$-sized cluster state with a fidelity of $F > 0.5$ can be generated at 65-kHz repetition rate -- well beyond the current state of the art for proposals based on single photons \cite{Istrati2019}. The scheme can be applied directly to the other group-IV diamond colour centres such as tin vacancy and germanium vacancy, both having nuclear spin isotopes. The latter is host to a $\frac{9}{2}$-nuclear spin that could be used as a 3-qubit nuclear register to extend the spin-dimension to 4 rails \cite{Chekhovich2020}. Our proposal can also be adapted in a straightforward manner to other spin-photon interfaces, such as self-assembled QDs, silicon carbide, or rare-earth ions.

\section{Acknowledgements}

We acknowledge support from the ERC Advanced Grant PEDESTAL (884745), and the EU Quantum Flagship 2D-SIPC. Cathryn Michaels acknowledges support from the EPSRC DTP. Jes\'{u}s Arjona Mart\'{i}nez from the Winton Programme and EPSRC DTP, Romain Debroux from the Cambridge Gates Trust, Alexander Stramma from EPSRC/NQIT, Ryan Parker from the General Sir John Monash Foundation, and Dorian Gangloff from a St John's College Title A Fellowship.

\bibliographystyle{elsarticle-num}
\bibliography{references3}

\begin{thebibliography}{100}
\expandafter\ifx\csname url\endcsname\relax
  \def\url#1{\texttt{#1}}\fi
\expandafter\ifx\csname urlprefix\endcsname\relax\def\urlprefix{URL }\fi
\expandafter\ifx\csname href\endcsname\relax
  \def\href#1#2{#2} \def\path#1{#1}\fi

\bibitem{Aspect1981}
A.~Aspect, P.~Grangier, G.~Roger, {Experimental Tests of Realistic Local
  Theories via Bell's Theorem}, Phys. Rev. Lett. 47~(7) (1981) 460--463.
\newblock \href {https://doi.org/10.1103/PhysRevLett.47.460}
  {\path{doi:10.1103/PhysRevLett.47.460}}.

\bibitem{Ekert1991}
A.~K. Ekert, {Quantum cryptography based on Bell's theorem}, Phys. Rev. Lett.
  67~(6) (1991) 661--663.
\newblock \href {https://doi.org/10.1103/PhysRevLett.67.661}
  {\path{doi:10.1103/PhysRevLett.67.661}}.

\bibitem{Bouwmeester1997}
D.~Bouwmeester, J.-W. Pan, K.~Mattle, M.~Eibl, H.~Weinfurter, A.~Zeilinger,
  {Experimental quantum teleportation}, Nature 390~(6660) (1997) 575--579.
\newblock \href {https://doi.org/10.1038/37539} {\path{doi:10.1038/37539}}.

\bibitem{Raussendorf2001a}
R.~Raussendorf, H.~J. Briegel, {A One-Way Quantum Computer}, Phys. Rev. Lett.
  86~(22) (2001) 5188--5191.
\newblock \href {https://doi.org/10.1103/physrevlett.86.5188}
  {\path{doi:10.1103/physrevlett.86.5188}}.

\bibitem{Raussendorf2003}
R.~Raussendorf, D.~E. Browne, H.~J. Briegel, {Measurement-based quantum
  computation on cluster states}, Phys. Rev. A 68~(2) (2003) 022312.
\newblock \href {https://doi.org/10.1103/PhysRevA.68.022312}
  {\path{doi:10.1103/PhysRevA.68.022312}}.

\bibitem{Briegel2009a}
H.~J. Briegel, D.~E. Browne, W.~Dür, R.~Raussendorf, M.~V. den Nest,
  Measurement-based quantum computation, Nat. Phys. 5~(1) (2009) 19--26.
\newblock \href {https://doi.org/10.1038/nphys1157}
  {\path{doi:10.1038/nphys1157}}.

\bibitem{Kimble2008}
H.~J. Kimble, {The quantum internet}, Nature 453~(7198) (2008) 1023--1030.
\newblock \href {https://doi.org/10.1038/nature07127}
  {\path{doi:10.1038/nature07127}}.

\bibitem{Ladd2010}
T.~D. Ladd, F.~Jelezko, R.~Laflamme, Y.~Nakamura, C.~Monroe, J.~L. O'Brien,
  {Quantum computers.}, Nature 464~(7285) (2010) 45--53.
\newblock \href {https://doi.org/10.1038/nature08812}
  {\path{doi:10.1038/nature08812}}.

\bibitem{Gisin2002}
N.~Gisin, G.~Ribordy, W.~Tittel, H.~Zbinden, {Quantum cryptography}, Rev. Mod.
  Phys. 74~(1) (2002) 145--195.
\newblock \href {https://doi.org/10.1103/RevModPhys.74.145}
  {\path{doi:10.1103/RevModPhys.74.145}}.

\bibitem{Zhong2020}
H.-S. Zhong, H.~Wang, Y.-H. Deng, M.-C. Chen, L.-C. Peng, Y.-H. Luo, J.~Qin,
  D.~Wu, X.~Ding, Y.~Hu, P.~Hu, X.-Y. Yang, W.-J. Zhang, H.~Li, Y.~Li,
  X.~Jiang, L.~Gan, G.~Yang, L.~You, Z.~Wang, L.~Li, N.-L. Liu, C.-Y. Lu, J.-W.
  Pan, Quantum computational advantage using photons, Science 370~(6523) (2020)
  1460--1463.
\newblock \href {https://doi.org/10.1126/science.abe8770}
  {\path{doi:10.1126/science.abe8770}}.

\bibitem{Xu2020}
F.~Xu, X.~Ma, Q.~Zhang, H.-K. Lo, J.-W. Pan, {Secure quantum key distribution
  with realistic devices}, Rev. Mod. Phys. 92~(2) (2020) 025002.
\newblock \href {https://doi.org/10.1103/RevModPhys.92.025002}
  {\path{doi:10.1103/RevModPhys.92.025002}}.

\bibitem{Briegel2001}
H.~J. Briegel, R.~Raussendorf, {Persistent entanglement in arrays of
  interacting particles}, Phys. Rev. Lett. 86~(5) (2001) 910--913.
\newblock \href {https://doi.org/10.1103/PhysRevLett.86.910}
  {\path{doi:10.1103/PhysRevLett.86.910}}.

\bibitem{Varnava2008}
M.~Varnava, D.~E. Browne, T.~Rudolph, {How Good Must Single Photon Sources and
  Detectors Be for Efficient Linear Optical Quantum Computation?}, Phys. Rev.
  Lett. 100~(6) (2008) 060502.
\newblock \href {https://doi.org/10.1103/PhysRevLett.100.060502}
  {\path{doi:10.1103/PhysRevLett.100.060502}}.

\bibitem{Zwerger2016}
M.~Zwerger, H.~J. Briegel, W.~D{\"{u}}r, {Measurement-based quantum
  communication}, Appl. Phys. B 122~(3) (2016) 50.
\newblock \href {https://doi.org/10.1007/s00340-015-6285-8}
  {\path{doi:10.1007/s00340-015-6285-8}}.

\bibitem{Azuma2015All-photonicRepeaters}
K.~Azuma, K.~Tamaki, H.-K. Lo, {All-photonic quantum repeaters}, Nat. Commun.
  6~(1) (2015) 6787.
\newblock \href {https://doi.org/10.1038/ncomms7787}
  {\path{doi:10.1038/ncomms7787}}.

\bibitem{Grice2011ArbitrarilyElements}
W.~P. Grice, {Arbitrarily complete Bell-state measurement using only linear
  optical elements}, Phys. Rev. A 84~(4) (2011) 042331.
\newblock \href {https://doi.org/10.1103/PhysRevA.84.042331}
  {\path{doi:10.1103/PhysRevA.84.042331}}.

\bibitem{Kilmer2019BoostingDetectors}
T.~Kilmer, S.~Guha, {Boosting linear-optical Bell measurement success
  probability with predetection squeezing and imperfect photon-number-resolving
  detectors}, Phys. Rev. A 99~(3) (2019) 032302.
\newblock \href {https://doi.org/10.1103/PhysRevA.99.032302}
  {\path{doi:10.1103/PhysRevA.99.032302}}.

\bibitem{Ewert20143/4Ancillae}
F.~Ewert, P.~van Loock, {3/4-Efficient Bell Measurement with Passive Linear
  Optics and Unentangled Ancillae}, Phys. Rev. Lett. 113~(14) (2014) 140403.
\newblock \href {https://doi.org/10.1103/PhysRevLett.113.140403}
  {\path{doi:10.1103/PhysRevLett.113.140403}}.

\bibitem{Browne2005Resource-EfficientComputation}
D.~E. Browne, T.~Rudolph, {Resource-Efficient Linear Optical Quantum
  Computation}, Phys. Rev. Lett. 95~(1) (2005) 010501.
\newblock \href {https://doi.org/10.1103/PhysRevLett.95.010501}
  {\path{doi:10.1103/PhysRevLett.95.010501}}.

\bibitem{Zhao2004ExperimentalTeleportation}
Z.~Zhao, Y.-A. Chen, A.-N. Zhang, T.~Yang, H.~J. Briegel, J.-W. Pan,
  {Experimental demonstration of five-photon entanglement and open-destination
  teleportation}, Nature 430~(6995) (2004) 54--58.
\newblock \href {https://doi.org/10.1038/nature02643}
  {\path{doi:10.1038/nature02643}}.

\bibitem{Gao2010ExperimentalState}
W.~B. Gao, C.~Y. Lu, X.~C. Yao, P.~Xu, O.~G{\"{u}}hne, A.~Goebel, Y.~A. Chen,
  C.~Z. Peng, Z.~B. Chen, J.~W. Pan, {Experimental demonstration of a
  hyper-entangled ten-qubit Schr{\"{o}}dinger cat state}, Nat. Phys. 6~(5)
  (2010) 331--335.
\newblock \href {https://doi.org/10.1038/nphys1603}
  {\path{doi:10.1038/nphys1603}}.

\bibitem{Wang2016ExperimentalEntanglement}
X.-L. Wang, L.-K. Chen, W.~Li, H.-L. Huang, C.~Liu, C.~Chen, Y.-H. Luo, Z.-E.
  Su, D.~Wu, Z.-D. Li, H.~Lu, Y.~Hu, X.~Jiang, C.-Z. Peng, L.~Li, N.-L. Liu,
  Y.-A. Chen, C.-Y. Lu, J.-W. Pan, {Experimental Ten-Photon Entanglement},
  Phys. Rev. Lett. 117~(21) (2016) 210502.
\newblock \href {https://doi.org/10.1103/PhysRevLett.117.210502}
  {\path{doi:10.1103/PhysRevLett.117.210502}}.

\bibitem{Istrati2019}
D.~Istrati, Y.~Pilnyak, J.~C. Loredo, C.~Ant{\'{o}}n, N.~Somaschi, P.~Hilaire,
  H.~Ollivier, M.~Esmann, L.~Cohen, L.~Vidro, C.~Millet, A.~Lema{\^{i}}tre,
  I.~Sagnes, A.~Harouri, L.~Lanco, P.~Senellart, H.~S. Eisenberg, {Sequential
  generation of linear cluster states from a single photon emitter}, Nat.
  Commun. 11~(1) (2020) 5501.
\newblock \href {https://doi.org/10.1038/s41467-020-19341-4}
  {\path{doi:10.1038/s41467-020-19341-4}}.

\bibitem{Asavanant2019}
W.~Asavanant, Y.~Shiozawa, S.~Yokoyama, B.~Charoensombutamon, H.~Emura, R.~N.
  Alexander, S.~Takeda, J.-i. Yoshikawa, N.~C. Menicucci, H.~Yonezawa,
  A.~Furusawa, {Generation of time-domain-multiplexed two-dimensional cluster
  state}, Science 366~(6463) (2019) 373--376.
\newblock \href {https://doi.org/10.1126/science.aay2645}
  {\path{doi:10.1126/science.aay2645}}.

\bibitem{Lindner2009a}
N.~H. Lindner, T.~Rudolph, {Proposal for Pulsed On-Demand Sources of Photonic
  Cluster State Strings}, Phys. Rev. Lett. 103~(11) (2009) 113602.
\newblock \href {https://doi.org/10.1103/PhysRevLett.103.113602}
  {\path{doi:10.1103/PhysRevLett.103.113602}}.

\bibitem{Schwartz2016}
I.~Schwartz, D.~Cogan, E.~R. Schmidgall, Y.~Don, L.~Gantz, O.~Kenneth, N.~H.
  Lindner, D.~Gershoni, {Deterministic generation of a cluster state of
  entangled photons}, Science 354~(6311) (2016) 434--437.
\newblock \href {https://doi.org/10.1126/science.aah4758}
  {\path{doi:10.1126/science.aah4758}}.

\bibitem{Gontagonta2009}
D.~Gonţa, T.~Radtke, S.~Fritzsche, {Generation of two-dimensional cluster
  states by using high-finesse bimodal cavities}, Phys. Rev. A 79~(6) (2009)
  062319.
\newblock \href {https://doi.org/10.1103/PhysRevA.79.062319}
  {\path{doi:10.1103/PhysRevA.79.062319}}.

\bibitem{Economou2010b}
S.~E. Economou, N.~Lindner, T.~Rudolph, {Optically Generated 2-Dimensional
  Photonic Cluster State from Coupled Quantum Dots}, Phys. Rev. Lett. 105~(9)
  (2010) 093601.
\newblock \href {https://doi.org/10.1103/PhysRevLett.105.093601}
  {\path{doi:10.1103/PhysRevLett.105.093601}}.

\bibitem{Mantri2017}
A.~Mantri, T.~F. Demarie, J.~F. Fitzsimons, {Universality of quantum
  computation with cluster states and (X, Y)-plane measurements}, Sci. Rep.
  7~(1) (2017) 42861.
\newblock \href {https://doi.org/10.1038/srep42861}
  {\path{doi:10.1038/srep42861}}.

\bibitem{Gimeno-Segovia2018}
M.~Gimeno-Segovia, T.~Rudolph, S.~E. Economou, {Deterministic Generation of
  Large-Scale Entangled Photonic Cluster State from Interacting Solid State
  Emitters}, Phys. Rev. Lett. 123~(7) (2019) 070501.
\newblock \href {https://doi.org/10.1103/PhysRevLett.123.070501}
  {\path{doi:10.1103/PhysRevLett.123.070501}}.

\bibitem{Russo2019}
A.~Russo, E.~Barnes, S.~E. Economou, {Generation of arbitrary all-photonic
  graph states from quantum emitters}, New J. Phys. 21~(5) (2019) 055002.
\newblock \href {https://doi.org/10.1088/1367-2630/ab193d}
  {\path{doi:10.1088/1367-2630/ab193d}}.

\bibitem{Russo2018PhotonicCommunications}
A.~Russo, E.~Barnes, S.~E. Economou, {Photonic graph state generation from
  quantum dots and color centers for quantum communications}, Phys. Rev. B
  98~(8) (2018) 085303.
\newblock \href {https://doi.org/10.1103/PhysRevB.98.085303}
  {\path{doi:10.1103/PhysRevB.98.085303}}.

\bibitem{Buterakos2017DeterministicEmitters}
D.~Buterakos, E.~Barnes, S.~E. Economou, {Deterministic Generation of
  All-Photonic Quantum Repeaters from Solid-State Emitters}, Phys. Rev. X 7~(4)
  (2017) 041023.
\newblock \href {https://doi.org/10.1103/PhysRevX.7.041023}
  {\path{doi:10.1103/PhysRevX.7.041023}}.

\bibitem{waldherr2014quantum}
G.~Waldherr, Y.~Wang, S.~Zaiser, M.~Jamali, T.~Schulte-Herbr{\"u}ggen, H.~Abe,
  T.~Ohshima, J.~Isoya, J.~F. Du, P.~Neumann, J.~Wrachtrup, Quantum error
  correction in a solid-state hybrid spin register, Nature 506~(7487) (2014)
  204--207.
\newblock \href {https://doi.org/10.1038/nature12919}
  {\path{doi:10.1038/nature12919}}.

\bibitem{gangloff2019nuclearensemble}
D.~A. Gangloff, G.~{\'{E}}thier-Majcher, C.~Lang, E.~V. Denning, J.~H. Bodey,
  D.~M. Jackson, E.~Clarke, M.~Hugues, C.~{Le Gall}, M.~Atat{\"{u}}re, {Quantum
  interface of an electron and a nuclear ensemble}, Science 364~(6435) (2019)
  62--66.
\newblock \href {https://doi.org/10.1126/science.aaw2906}
  {\path{doi:10.1126/science.aaw2906}}.

\bibitem{metsch2019nuclearsiv}
M.~H. Metsch, K.~Senkalla, B.~Tratzmiller, J.~Scheuer, M.~Kern, J.~Achard,
  A.~Tallaire, M.~B. Plenio, P.~Siyushev, F.~Jelezko, {Initialization and
  Readout of Nuclear Spins via a Negatively Charged Silicon-Vacancy Center in
  Diamond}, Phys. Rev. Lett. 122~(19) (2019) 190503.
\newblock \href {https://doi.org/10.1103/PhysRevLett.122.190503}
  {\path{doi:10.1103/PhysRevLett.122.190503}}.

\bibitem{Atature2018}
M.~Atat{\"{u}}re, D.~Englund, N.~Vamivakas, S.-Y. Lee, J.~Wrachtrup, {Material
  platforms for spin-based photonic quantum technologies}, Nat. Rev. Mater.
  3~(5) (2018) 38--51.
\newblock \href {https://doi.org/10.1038/s41578-018-0008-9}
  {\path{doi:10.1038/s41578-018-0008-9}}.

\bibitem{janitz2020cavity}
E.~Janitz, M.~K. Bhaskar, L.~Childress, {Cavity quantum electrodynamics with
  color centers in diamond}, Optica 7~(10) (2020) 1232.
\newblock \href {https://doi.org/10.1364/OPTICA.398628}
  {\path{doi:10.1364/OPTICA.398628}}.

\bibitem{OBrien2009}
J.~L. O'Brien, A.~Furusawa, J.~Vu{\v{c}}kovi{\'{c}}, {Photonic quantum
  technologies}, Nat. Photonics 3~(12) (2009) 687--695.
\newblock \href {https://doi.org/10.1038/nphoton.2009.229}
  {\path{doi:10.1038/nphoton.2009.229}}.

\bibitem{Paillard2000}
M.~Paillard, X.~Marie, E.~Vanelle, T.~Amand, V.~K. Kalevich, A.~R. Kovsh, A.~E.
  Zhukov, V.~M. Ustinov, {Time-resolved photoluminescence in self-assembled
  InAs/GaAs quantum dots under strictly resonant excitation}, Appl. Phys. Lett.
  76~(1) (2000) 76--78.
\newblock \href {https://doi.org/10.1063/1.125661}
  {\path{doi:10.1063/1.125661}}.

\bibitem{najer2019gated}
D.~Najer, I.~S{\"{o}}llner, P.~Sekatski, V.~Dolique, M.~C. L{\"{o}}bl,
  D.~Riedel, R.~Schott, S.~Starosielec, S.~R. Valentin, A.~D. Wieck,
  N.~Sangouard, A.~Ludwig, R.~J. Warburton, {A gated quantum dot strongly
  coupled to an optical microcavity}, Nature 575~(7784) (2019) 622--627.
\newblock \href {https://doi.org/10.1038/s41586-019-1709-y}
  {\path{doi:10.1038/s41586-019-1709-y}}.

\bibitem{Senellart2017}
P.~Senellart, G.~Solomon, A.~White, {High-performance semiconductor quantum-dot
  single-photon sources}, Nat. Nanotechnol. 12~(11) (2017) 1026--1039.
\newblock \href {https://doi.org/10.1038/nnano.2017.218}
  {\path{doi:10.1038/nnano.2017.218}}.

\bibitem{peter2004phonon}
E.~Peter, J.~Hours, P.~Senellart, A.~Vasanelli, A.~Cavanna, J.~Bloch, J.~M.
  G{\'{e}}rard, {Phonon sidebands in exciton and biexciton emission from single
  GaAs quantum dots}, Phys. Rev. B 69~(4) (2004) 041307.
\newblock \href {https://doi.org/10.1103/PhysRevB.69.041307}
  {\path{doi:10.1103/PhysRevB.69.041307}}.

\bibitem{matthiesen2013phase}
C.~Matthiesen, M.~Geller, C.~H.~H. Schulte, C.~{Le Gall}, J.~Hansom, Z.~Li,
  M.~Hugues, E.~Clarke, M.~Atat{\"{u}}re, {Phase-locked indistinguishable
  photons with synthesized waveforms from a solid-state source}, Nat. Commun.
  4~(1) (2013) 1600.
\newblock \href {https://doi.org/10.1038/ncomms2601}
  {\path{doi:10.1038/ncomms2601}}.

\bibitem{konthasinghe2012coherent}
K.~Konthasinghe, J.~Walker, M.~Peiris, C.~K. Shih, Y.~Yu, M.~F. Li, J.~F. He,
  L.~J. Wang, H.~Q. Ni, Z.~C. Niu, A.~Muller, {Coherent versus incoherent light
  scattering from a quantum dot}, Phys. Rev. B 85~(23) (2012) 235315.
\newblock \href {https://doi.org/10.1103/PhysRevB.85.235315}
  {\path{doi:10.1103/PhysRevB.85.235315}}.

\bibitem{Bechtold2015}
A.~Bechtold, D.~Rauch, F.~Li, T.~Simmet, P.-L. Ardelt, A.~Regler,
  K.~M{\"{u}}ller, N.~A. Sinitsyn, J.~J. Finley, {Three-stage decoherence
  dynamics of an electron spin qubit in an optically active quantum dot}, Nat.
  Phys. 11~(12) (2015) 1005--1008.
\newblock \href {https://doi.org/10.1038/nphys3470}
  {\path{doi:10.1038/nphys3470}}.

\bibitem{Stockill2016}
R.~Stockill, C.~{Le Gall}, C.~Matthiesen, L.~Huthmacher, E.~Clarke, M.~Hugues,
  M.~Atat{\"{u}}re, {Quantum dot spin coherence governed by a strained nuclear
  environment}, Nat. Commun. 7~(1) (2016) 12745.
\newblock \href {https://doi.org/10.1038/ncomms12745}
  {\path{doi:10.1038/ncomms12745}}.

\bibitem{Hogele2012a}
A.~H{\"{o}}gele, M.~Kroner, C.~Latta, M.~Claassen, I.~Carusotto, C.~Bulutay,
  A.~Imamoglu, {Dynamic Nuclear Spin Polarization in the Resonant Laser
  Excitation of an InGaAs Quantum Dot}, Phys. Rev. Lett. 108~(19) (2012)
  197403.
\newblock \href {https://doi.org/10.1103/PhysRevLett.108.197403}
  {\path{doi:10.1103/PhysRevLett.108.197403}}.

\bibitem{christle2017isolated}
D.~J. Christle, P.~V. Klimov, C.~F. de~las Casas, K.~Sz{\'{a}}sz,
  V.~Iv{\'{a}}dy, V.~Jokubavicius, J.~{Ul Hassan}, M.~Syv{\"{a}}j{\"{a}}rvi,
  W.~F. Koehl, T.~Ohshima, N.~T. Son, E.~Janz{\'{e}}n, {\'{A}}.~Gali, D.~D.
  Awschalom, {Isolated Spin Qubits in SiC with a High-Fidelity Infrared
  Spin-to-Photon Interface}, Phys. Rev. X 7~(2) (2017) 021046.
\newblock \href {https://doi.org/10.1103/PhysRevX.7.021046}
  {\path{doi:10.1103/PhysRevX.7.021046}}.

\bibitem{calusine2014silicon}
G.~Calusine, A.~Politi, D.~D. Awschalom, Silicon carbide photonic crystal
  cavities with integrated color centers, Appl. Phys. Lett. 105~(1) (2014)
  011123.
\newblock \href {https://doi.org/10.1063/1.4890083}
  {\path{doi:10.1063/1.4890083}}.

\bibitem{Bourassa2020}
A.~Bourassa, C.~P. Anderson, K.~C. Miao, M.~Onizhuk, H.~Ma, A.~L. Crook,
  H.~Abe, J.~Ul-Hassan, T.~Ohshima, N.~T. Son, G.~Galli, D.~D. Awschalom,
  {Entanglement and control of single nuclear spins in isotopically engineered
  silicon carbide}, Nat. Mater. 19~(12) (2020) 1319--1325.
\newblock \href {https://doi.org/10.1038/s41563-020-00802-6}
  {\path{doi:10.1038/s41563-020-00802-6}}.

\bibitem{Spindlberger2019}
L.~Spindlberger, A.~Cs{\'{o}}r{\'{e}}, G.~Thiering, S.~Putz, R.~Karhu, J.~U.
  Hassan, N.~T. Son, T.~Fromherz, A.~Gali, M.~Trupke, {Optical Properties of
  Vanadium in 4 H Silicon Carbide for Quantum Technology}, Phys. Rev. Applied
  12~(1) (2019) 014015.
\newblock \href {https://doi.org/10.1103/PhysRevApplied.12.014015}
  {\path{doi:10.1103/PhysRevApplied.12.014015}}.

\bibitem{wolfowicz2020vanadium}
G.~Wolfowicz, C.~P. Anderson, B.~Diler, O.~G. Poluektov, F.~J. Heremans, D.~D.
  Awschalom, {Vanadium spin qubits as telecom quantum emitters in silicon
  carbide}, Sci. Adv. 6~(18) (2020) eaaz1192.
\newblock \href {https://doi.org/10.1126/sciadv.aaz1192}
  {\path{doi:10.1126/sciadv.aaz1192}}.

\bibitem{manson2006nvstructure}
N.~B. Manson, J.~P. Harrison, M.~J. Sellars, {Nitrogen-vacancy center in
  diamond: Model of the electronic structure and associated dynamics}, Phys.
  Rev. B 74~(10) (2006) 104303.
\newblock \href {https://doi.org/10.1103/PhysRevB.74.104303}
  {\path{doi:10.1103/PhysRevB.74.104303}}.

\bibitem{riedel2017deterministic}
D.~Riedel, I.~S{\"{o}}llner, B.~J. Shields, S.~Starosielec, P.~Appel, E.~Neu,
  P.~Maletinsky, R.~J. Warburton, {Deterministic Enhancement of Coherent Photon
  Generation from a Nitrogen-Vacancy Center in Ultrapure Diamond}, Phys. Rev. X
  7~(3) (2017) 031040.
\newblock \href {https://doi.org/10.1103/PhysRevX.7.031040}
  {\path{doi:10.1103/PhysRevX.7.031040}}.

\bibitem{Berthel2015}
M.~Berthel, O.~Mollet, G.~Dantelle, T.~Gacoin, S.~Huant, A.~Drezet,
  {Photophysics of single nitrogen-vacancy centers in diamond nanocrystals},
  Phys. Rev. B 91~(3) (2015) 035308.
\newblock \href {https://doi.org/10.1103/PhysRevB.91.035308}
  {\path{doi:10.1103/PhysRevB.91.035308}}.

\bibitem{Patel2016}
R.~N. Patel, T.~Schr{\"{o}}der, N.~Wan, L.~Li, S.~L. Mouradian, E.~H. Chen,
  D.~R. Englund, {Efficient photon coupling from a diamond nitrogen vacancy
  center by integration with silica fiber}, Light Sci. Appl. 5~(2) (2016)
  e16032--e16032.
\newblock \href {https://doi.org/10.1038/lsa.2016.32}
  {\path{doi:10.1038/lsa.2016.32}}.

\bibitem{aharonovich2011diamond}
I.~Aharonovich, S.~Castelletto, D.~A. Simpson, C.-H. Su, A.~D. Greentree,
  S.~Prawer, {Diamond-based single-photon emitters}, Reports Prog. Phys. 74~(7)
  (2011) 076501.
\newblock \href {https://doi.org/10.1088/0034-4885/74/7/076501}
  {\path{doi:10.1088/0034-4885/74/7/076501}}.

\bibitem{humphreys2017deterministicentanglement}
P.~C. Humphreys, N.~Kalb, J.~P. Morits, R.~N. Schouten, R.~F. Vermeulen, D.~J.
  Twitchen, M.~Markham, R.~Hanson, {Deterministic delivery of remote
  entanglement on a quantum network}, Nature 558~(7709) (2018) 268--273.
\newblock \href {https://doi.org/10.1038/s41586-018-0200-5}
  {\path{doi:10.1038/s41586-018-0200-5}}.

\bibitem{Pfaff2013}
W.~Pfaff, T.~H. Taminiau, L.~Robledo, H.~Bernien, M.~Markham, D.~J. Twitchen,
  R.~Hanson, {Demonstration of entanglement-by-measurement of solid-state
  qubits}, Nat. Phys. 9~(1) (2013) 29--33.
\newblock \href {https://doi.org/10.1038/nphys2444}
  {\path{doi:10.1038/nphys2444}}.

\bibitem{becker2018siv}
J.~N. Becker, B.~Pingault, D.~Gro{\ss}, M.~G{\"{u}}ndoğan, N.~Kukharchyk,
  M.~Markham, A.~Edmonds, M.~Atat{\"{u}}re, P.~Bushev, C.~Becher, {All-Optical
  Control of the Silicon-Vacancy Spin in Diamond at Millikelvin Temperatures},
  Phys. Rev. Lett. 120~(5) (2018) 053603.
\newblock \href {https://doi.org/10.1103/PhysRevLett.120.053603}
  {\path{doi:10.1103/PhysRevLett.120.053603}}.

\bibitem{bhaskar2020memoryenhanced}
M.~K. Bhaskar, R.~Riedinger, B.~Machielse, D.~S. Levonian, C.~T. Nguyen, E.~N.
  Knall, H.~Park, D.~Englund, M.~Lon{\v c}ar, D.~D. Sukachev, M.~D. Lukin,
  {Experimental demonstration of memory-enhanced quantum communication}, Nature
  580~(7801) (2020) 60--64.
\newblock \href {https://doi.org/10.1038/s41586-020-2103-5}
  {\path{doi:10.1038/s41586-020-2103-5}}.

\bibitem{sukachev2017siv}
D.~D. Sukachev, A.~Sipahigil, C.~T. Nguyen, M.~K. Bhaskar, R.~E. Evans,
  F.~Jelezko, M.~D. Lukin, {Silicon-Vacancy Spin Qubit in Diamond: A Quantum
  Memory Exceeding 10 ms with Single-Shot State Readout}, Phys. Rev. Lett.
  119~(22) (2017) 223602.
\newblock \href {https://doi.org/10.1103/PhysRevLett.119.223602}
  {\path{doi:10.1103/PhysRevLett.119.223602}}.

\bibitem{Neu2011FluorescenceIridium}
E.~Neu, M.~Fischer, S.~Gsell, M.~Schreck, C.~Becher, {Fluorescence and
  polarization spectroscopy of single silicon vacancy centers in
  heteroepitaxial nanodiamonds on iridium}, Phys. Rev. B 84~(20) (2011) 205211.
\newblock \href {https://doi.org/10.1103/PhysRevB.84.205211}
  {\path{doi:10.1103/PhysRevB.84.205211}}.

\bibitem{Neu2011}
E.~Neu, D.~Steinmetz, J.~Riedrich-M{\"{o}}ller, S.~Gsell, M.~Fischer,
  M.~Schreck, C.~Becher, {Single photon emission from silicon-vacancy colour
  centres in chemical vapour deposition nano-diamonds on iridium}, New J. Phys.
  13~(2) (2011) 025012.
\newblock \href {https://doi.org/10.1088/1367-2630/13/2/025012}
  {\path{doi:10.1088/1367-2630/13/2/025012}}.

\bibitem{Pingault2017CoherentDiamond}
B.~Pingault, D.-D. Jarausch, C.~Hepp, L.~Klintberg, J.~N. Becker, M.~Markham,
  C.~Becher, M.~Atat{\"{u}}re, {Coherent control of the silicon-vacancy spin in
  diamond}, Nat. Commun. 8~(1) (2017) 15579.
\newblock \href {https://doi.org/10.1038/ncomms15579}
  {\path{doi:10.1038/ncomms15579}}.

\bibitem{Edmonds2008ElectronDiamond}
A.~M. Edmonds, M.~E. Newton, P.~M. Martineau, D.~J. Twitchen, S.~D. Williams,
  {Electron paramagnetic resonance studies of silicon-related defects in
  diamond}, Phys. Rev. B 77~(24) (2008) 245205.
\newblock \href {https://doi.org/10.1103/PhysRevB.77.245205}
  {\path{doi:10.1103/PhysRevB.77.245205}}.

\bibitem{Iwasaki2015a}
T.~Iwasaki, F.~Ishibashi, Y.~Miyamoto, Y.~Doi, S.~Kobayashi, T.~Miyazaki,
  K.~Tahara, K.~D. Jahnke, L.~J. Rogers, B.~Naydenov, F.~Jelezko, S.~Yamasaki,
  S.~Nagamachi, T.~Inubushi, N.~Mizuochi, M.~Hatano, {Germanium-Vacancy Single
  Color Centers in Diamond}, Sci. Rep. 5~(1) (2015) 12882.
\newblock \href {https://doi.org/10.1038/srep12882}
  {\path{doi:10.1038/srep12882}}.

\bibitem{bhaskar2017quantum}
M.~K. Bhaskar, D.~D. Sukachev, A.~Sipahigil, R.~E. Evans, M.~J. Burek, C.~T.
  Nguyen, L.~J. Rogers, P.~Siyushev, M.~H. Metsch, H.~Park, F.~Jelezko,
  M.~Lon\ifmmode~\check{c}\else \v{c}\fi{}ar, M.~D. Lukin, Quantum nonlinear
  optics with a germanium-vacancy color center in a nanoscale diamond
  waveguide, Phys. Rev. Lett. 118 (2017) 223603.
\newblock \href {https://doi.org/10.1103/PhysRevLett.118.223603}
  {\path{doi:10.1103/PhysRevLett.118.223603}}.

\bibitem{Palyanov2015}
Y.~N. Palyanov, I.~N. Kupriyanov, Y.~M. Borzdov, N.~V. Surovtsev, {Germanium: a
  new catalyst for diamond synthesis and a new optically active impurity in
  diamond}, Sci. Rep. 5~(1) (2015) 14789.
\newblock \href {https://doi.org/10.1038/srep14789}
  {\path{doi:10.1038/srep14789}}.

\bibitem{trusheim2020transformlimitted}
M.~E. Trusheim, B.~Pingault, N.~H. Wan, M.~G{\"{u}}ndoğan, L.~{De Santis},
  R.~Debroux, D.~Gangloff, C.~Purser, K.~C. Chen, M.~Walsh, J.~J. Rose, J.~N.
  Becker, B.~Lienhard, E.~Bersin, I.~Paradeisanos, G.~Wang, D.~Lyzwa, A.~R.-P.
  Montblanch, G.~Malladi, H.~Bakhru, A.~C. Ferrari, I.~A. Walmsley,
  M.~Atat{\"{u}}re, D.~Englund, {Transform-Limited Photons From a Coherent
  Tin-Vacancy Spin in Diamond}, Phys. Rev. Lett. 124~(2) (2020) 023602.
\newblock \href {https://doi.org/10.1103/PhysRevLett.124.023602}
  {\path{doi:10.1103/PhysRevLett.124.023602}}.

\bibitem{rugar2021quantum}
A.~E. Rugar, S.~Aghaeimeibodi, D.~Riedel, C.~Dory, H.~Lu, P.~J. McQuade, Z.-X.
  Shen, N.~A. Melosh, J.~Vu{\v{c}}kovi{\'{c}}, {Quantum Photonic Interface for
  Tin-Vacancy Centers in Diamond}, Phys. Rev. X 11~(3) (2021) 031021.
\newblock \href {https://doi.org/10.1103/PhysRevX.11.031021}
  {\path{doi:10.1103/PhysRevX.11.031021}}.

\bibitem{Iwasaki2017}
T.~Iwasaki, Y.~Miyamoto, T.~Taniguchi, P.~Siyushev, M.~H. Metsch, F.~Jelezko,
  M.~Hatano, {Tin-Vacancy Quantum Emitters in Diamond}, Phys. Rev. Lett.
  119~(25) (2017) 253601.
\newblock \href {https://doi.org/10.1103/PhysRevLett.119.253601}
  {\path{doi:10.1103/PhysRevLett.119.253601}}.

\bibitem{gorlitz2020spectroscopic}
J.~G{\"{o}}rlitz, D.~Herrmann, G.~Thiering, P.~Fuchs, M.~Gandil, T.~Iwasaki,
  T.~Taniguchi, M.~Kieschnick, J.~Meijer, M.~Hatano, A.~Gali, C.~Becher,
  {Spectroscopic investigations of negatively charged tin-vacancy centres in
  diamond}, New J. Phys. 22~(1) (2020) 013048.
\newblock \href {https://doi.org/10.1088/1367-2630/ab6631}
  {\path{doi:10.1088/1367-2630/ab6631}}.

\bibitem{Debroux2021}
R.~Debroux, C.~P. Michaels, C.~M. Purser, N.~Wan, M.~E. Trusheim, J.~A.
  Mart{\'{i}}nez, R.~A. Parker, A.~M. Stramma, K.~C. Chen, L.~de~Santis, E.~M.
  Alexeev, A.~C. Ferrari, D.~Englund, D.~A. Gangloff, M.~Atat{\"{u}}re,
  {Quantum control of the tin-vacancy spin qubit in diamond},
  \href{http://arxiv.org/abs/2106.00723}{arXiv:2106.00723} (2021).

\bibitem{Tomm2021}
N.~Tomm, A.~Javadi, N.~O. Antoniadis, D.~Najer, M.~C. L{\"{o}}bl, A.~R. Korsch,
  R.~Schott, S.~R. Valentin, A.~D. Wieck, A.~Ludwig, R.~J. Warburton, {A bright
  and fast source of coherent single photons}, Nat. Nanotechnol. 16~(4) (2021)
  399--403.
\newblock \href {https://doi.org/10.1038/s41565-020-00831-x}
  {\path{doi:10.1038/s41565-020-00831-x}}.

\bibitem{Kim2011UltrafastSpins}
D.~Kim, S.~G. Carter, A.~Greilich, A.~S. Bracker, D.~Gammon, {Ultrafast optical
  control of entanglement between two quantum-dot spins}, Nat. Phys. 7~(3)
  (2011) 223--229.
\newblock \href {https://doi.org/10.1038/nphys1863}
  {\path{doi:10.1038/nphys1863}}.

\bibitem{Ding2019CoherentInterface}
D.~Ding, M.~H. Appel, A.~Javadi, X.~Zhou, M.~C. L{\"{o}}bl, I.~S{\"{o}}llner,
  R.~Schott, C.~Papon, T.~Pregnolato, L.~Midolo, A.~D. Wieck, A.~Ludwig, R.~J.
  Warburton, T.~Schr{\"{o}}der, P.~Lodahl, {Coherent Optical Control of a
  Quantum-Dot Spin-Qubit in a Waveguide-Based Spin-Photon Interface}, Phys.
  Rev. Applied 11~(3) (2019) 031002.
\newblock \href {https://doi.org/10.1103/PhysRevApplied.11.031002}
  {\path{doi:10.1103/PhysRevApplied.11.031002}}.

\bibitem{bodey2019spinlocking}
J.~H. Bodey, R.~Stockill, E.~V. Denning, D.~A. Gangloff,
  G.~{\'{E}}thier-Majcher, D.~M. Jackson, E.~Clarke, M.~Hugues, C.~L. Gall,
  M.~Atat{\"{u}}re, {Optical spin locking of a solid-state qubit}, npj Quantum
  Inf. 5~(1) (2019) 95.
\newblock \href {https://doi.org/10.1038/s41534-019-0206-3}
  {\path{doi:10.1038/s41534-019-0206-3}}.

\bibitem{Denning2019}
E.~V. Denning, D.~A. Gangloff, M.~Atat{\"{u}}re, J.~M{\o}rk, C.~{Le Gall},
  {Collective Quantum Memory Activated by a Driven Central Spin}, Phys. Rev.
  Lett. 123~(14) (2019) 140502.
\newblock \href {https://doi.org/10.1103/PhysRevLett.123.140502}
  {\path{doi:10.1103/PhysRevLett.123.140502}}.

\bibitem{DeLasCasas2017}
C.~F. De~Las~Casas, D.~J. Christle, J.~Ul~Hassan, T.~Ohshima, N.~T. Son, D.~D.
  Awschalom, {Stark tuning and electrical charge state control of single
  divacancies in silicon carbide}, Appl. Phys. Lett. 111~(26) (2017) 262403.
\newblock \href {https://doi.org/10.1063/1.5004174}
  {\path{doi:10.1063/1.5004174}}.

\bibitem{Tran2016}
T.~T. Tran, K.~Bray, M.~J. Ford, M.~Toth, I.~Aharonovich, {Quantum emission
  from hexagonal boron nitride monolayers}, Nat. Nanotechnol. 11~(1) (2016)
  37--41.
\newblock \href {https://doi.org/10.1038/nnano.2015.242}
  {\path{doi:10.1038/nnano.2015.242}}.

\bibitem{zhong2017}
T.~Zhong, J.~M. Kindem, J.~Rochman, A.~Faraon, {Interfacing broadband photonic
  qubits to on-chip cavity-protected rare-earth ensembles}, Nat. Commun. 8~(1)
  (2017) 14107.
\newblock \href {https://doi.org/10.1038/ncomms14107}
  {\path{doi:10.1038/ncomms14107}}.

\bibitem{Aharonovich2011}
I.~Aharonovich, A.~D. Greentree, S.~Prawer, {Diamond photonics}, Nat. Photonics
  5~(7) (2011) 397--405.
\newblock \href {https://doi.org/10.1038/nphoton.2011.54}
  {\path{doi:10.1038/nphoton.2011.54}}.

\bibitem{Aharonovich2014a}
I.~Aharonovich, E.~Neu, {Diamond Nanophotonics}, Adv. Opt. Mater. 2~(10) (2014)
  911--928.
\newblock \href {https://doi.org/10.1002/adom.201400189}
  {\path{doi:10.1002/adom.201400189}}.

\bibitem{Aharonovich2016}
I.~Aharonovich, D.~Englund, M.~Toth, {Solid-state single-photon emitters}, Nat.
  Photonics 10~(10) (2016) 631--641.
\newblock \href {https://doi.org/10.1038/nphoton.2016.186}
  {\path{doi:10.1038/nphoton.2016.186}}.

\bibitem{fuchs2011quantummemory}
G.~D. Fuchs, G.~Burkard, P.~V. Klimov, D.~D. Awschalom, {A quantum memory
  intrinsic to single nitrogen-vacancy centres in diamond}, Nat. Phys. 7~(10)
  (2011) 789--793.
\newblock \href {https://doi.org/10.1038/nphys2026}
  {\path{doi:10.1038/nphys2026}}.

\bibitem{holzgrafe2019correctedreadout}
J.~Holzgrafe, J.~Beitner, D.~Kara, H.~S. Knowles, M.~Atat{\"{u}}re, {Error
  corrected spin-state readout in a nanodiamond}, npj Quantum Inf. 5~(1) (2019)
  13.
\newblock \href {https://doi.org/10.1038/s41534-019-0126-2}
  {\path{doi:10.1038/s41534-019-0126-2}}.

\bibitem{Togan2010}
E.~Togan, Y.~Chu, A.~S. Trifonov, L.~Jiang, J.~Maze, L.~Childress, M.~V.~G.
  Dutt, A.~S. S{\o}rensen, P.~R. Hemmer, A.~S. Zibrov, M.~D. Lukin, {Quantum
  entanglement between an optical photon and a solid-state spin qubit}, Nature
  466~(7307) (2010) 730--734.
\newblock \href {https://doi.org/10.1038/nature09256}
  {\path{doi:10.1038/nature09256}}.

\bibitem{Bradac2019a}
C.~Bradac, W.~Gao, J.~Forneris, M.~E. Trusheim, I.~Aharonovich, {Quantum
  nanophotonics with group IV defects in diamond}, Nat. Commun. 10~(1) (2019)
  5625.
\newblock \href {https://doi.org/10.1038/s41467-019-13332-w}
  {\path{doi:10.1038/s41467-019-13332-w}}.

\bibitem{Trusheim2019}
M.~E. Trusheim, N.~H. Wan, K.~C. Chen, C.~J. Ciccarino, J.~Flick,
  R.~Sundararaman, G.~Malladi, E.~Bersin, M.~Walsh, B.~Lienhard, H.~Bakhru,
  P.~Narang, D.~Englund, {Lead-related quantum emitters in diamond}, Phys. Rev.
  B 99~(7) (2019) 075430.
\newblock \href {https://doi.org/10.1103/PhysRevB.99.075430}
  {\path{doi:10.1103/PhysRevB.99.075430}}.

\bibitem{Wan2019}
N.~H. Wan, T.~J. Lu, K.~C. Chen, M.~P. Walsh, M.~E. Trusheim, L.~De~Santis,
  E.~A. Bersin, I.~B. Harris, S.~L. Mouradian, I.~R. Christen, E.~S. Bielejec,
  D.~Englund, {Large-scale integration of artificial atoms in hybrid photonic
  circuits}, Nature 583~(7815) (2020) 226--231.
\newblock \href {https://doi.org/10.1038/s41586-020-2441-3}
  {\path{doi:10.1038/s41586-020-2441-3}}.

\bibitem{kuruma2021}
K.~Kuruma, B.~Pingault, C.~Chia, D.~Renaud, P.~Hoffmann, S.~Iwamoto,
  C.~Ronning, M.~Lon{\v{c}}ar, Coupling of a single tin-vacancy center to a
  photonic crystal cavity in diamond, Applied Physics Letters 118~(23) (2021)
  230601.
\newblock \href {https://doi.org/10.1063/5.0051675}
  {\path{doi:10.1063/5.0051675}}.

\bibitem{fuchs2021}
P.~Fuchs, T.~Jung, M.~Kieschnick, J.~Meijer, C.~Becher, A cavity-based optical
  antenna for color centers in diamond, APL Photonics 6~(8) (2021) 086102.
\newblock \href {https://doi.org/10.1063/5.0057161}
  {\path{doi:10.1063/5.0057161}}.

\bibitem{Hepp2014}
C.~Hepp, T.~M{\"{u}}ller, V.~Waselowski, J.~N. Becker, B.~Pingault,
  H.~Sternschulte, D.~Steinm{\"{u}}ller-Nethl, A.~Gali, J.~R. Maze,
  M.~Atat{\"{u}}re, C.~Becher, {Electronic Structure of the Silicon Vacancy
  Color Center in Diamond}, Phys. Rev. Lett. 112~(3) (2014) 036405.
\newblock \href {https://doi.org/10.1103/PhysRevLett.112.036405}
  {\path{doi:10.1103/PhysRevLett.112.036405}}.

\bibitem{rogers2014siv}
L.~J. Rogers, K.~D. Jahnke, M.~W. Doherty, A.~Dietrich, L.~P. McGuinness,
  C.~M{\"{u}}ller, T.~Teraji, H.~Sumiya, J.~Isoya, N.~B. Manson, F.~Jelezko,
  {Electronic structure of the negatively charged silicon-vacancy center in
  diamond}, Phys. Rev. B 89~(23) (2014) 235101.
\newblock \href {https://doi.org/10.1103/PhysRevB.89.235101}
  {\path{doi:10.1103/PhysRevB.89.235101}}.

\bibitem{meesala2018strainengineering}
S.~Meesala, Y.-I. Sohn, B.~Pingault, L.~Shao, H.~A. Atikian, J.~Holzgrafe,
  M.~G{\"{u}}ndoğan, C.~Stavrakas, A.~Sipahigil, C.~Chia, R.~Evans, M.~J.
  Burek, M.~Zhang, L.~Wu, J.~L. Pacheco, J.~Abraham, E.~Bielejec, M.~D. Lukin,
  M.~Atat{\"{u}}re, M.~Lon{\v{c}}ar, {Strain engineering of the silicon-vacancy
  center in diamond}, Phys. Rev. B 97~(20) (2018) 205444.
\newblock \href {https://doi.org/10.1103/PhysRevB.97.205444}
  {\path{doi:10.1103/PhysRevB.97.205444}}.

\bibitem{Sohn2018a}
Y.-I. Sohn, S.~Meesala, B.~Pingault, H.~A. Atikian, J.~Holzgrafe,
  M.~G{\"{u}}ndoğan, C.~Stavrakas, M.~J. Stanley, A.~Sipahigil, J.~Choi,
  M.~Zhang, J.~L. Pacheco, J.~Abraham, E.~Bielejec, M.~D. Lukin,
  M.~Atat{\"{u}}re, M.~Lon{\v{c}}ar, {Controlling the coherence of a diamond
  spin qubit through its strain environment}, Nat. Commun. 9~(1) (2018) 2012.
\newblock \href {https://doi.org/10.1038/s41467-018-04340-3}
  {\path{doi:10.1038/s41467-018-04340-3}}.

\bibitem{Gali2013}
A.~Gali, J.~R. Maze, {Ab initio study of the split silicon-vacancy defect in
  diamond: Electronic structure and related properties}, Phys. Rev. B 88~(23)
  (2013) 235205.
\newblock \href {https://doi.org/10.1103/PhysRevB.88.235205}
  {\path{doi:10.1103/PhysRevB.88.235205}}.

\bibitem{Pingault2017}
B.~Pingault, {The silicon-vacancy centre in diamond for quantum information
  processing}, Ph.D. thesis, Cambridge (2017).
\newblock \href {https://doi.org/10.17863/CAM.15577}
  {\path{doi:10.17863/CAM.15577}}.

\bibitem{taminiau2014universalcontrol}
T.~H. Taminiau, J.~Cramer, T.~van~der Sar, V.~V. Dobrovitski, R.~Hanson,
  {Universal control and error correction in multi-qubit spin registers in
  diamond}, Nat. Nanotechnol. 9~(3) (2014) 171--176.
\newblock \href {https://doi.org/10.1038/nnano.2014.2}
  {\path{doi:10.1038/nnano.2014.2}}.

\bibitem{Schwartz2018}
I.~Schwartz, J.~Scheuer, B.~Tratzmiller, S.~M{\"{u}}ller, Q.~Chen, I.~Dhand,
  Z.-Y. Wang, C.~M{\"{u}}ller, B.~Naydenov, F.~Jelezko, M.~B. Plenio, {Robust
  optical polarization of nuclear spin baths using Hamiltonian engineering of
  nitrogen-vacancy center quantum dynamics}, Sci. Adv. 4~(8) (2018) eaat8978.
\newblock \href {https://doi.org/10.1126/sciadv.aat8978}
  {\path{doi:10.1126/sciadv.aat8978}}.

\bibitem{degreve2012}
K.~De~Greve, L.~Yu, P.~L. McMahon, J.~S. Pelc, C.~M. Natarajan, N.~Y. Kim,
  E.~Abe, S.~Maier, C.~Schneider, M.~Kamp, et~al., Quantum-dot spin--photon
  entanglement via frequency downconversion to telecom wavelength, Nature
  491~(7424) (2012) 421--425.
\newblock \href {https://doi.org/10.1038/nature11577}
  {\path{doi:10.1038/nature11577}}.

\bibitem{gao2012}
W.~Gao, P.~Fallahi, E.~Togan, J.~Miguel-S{\'a}nchez, A.~Imamoglu, Observation
  of entanglement between a quantum dot spin and a single photon, Nature
  491~(7424) (2012) 426--430.
\newblock \href {https://doi.org/10.1038/nature11573}
  {\path{doi:10.1038/nature11573}}.

\bibitem{Schaibley2013}
J.~R. Schaibley, A.~P. Burgers, G.~A. McCracken, L.-M. Duan, P.~R. Berman,
  D.~G. Steel, A.~S. Bracker, D.~Gammon, L.~J. Sham, {Demonstration of Quantum
  Entanglement between a Single Electron Spin Confined to an InAs Quantum Dot
  and a Photon}, Phys. Rev. Lett. 110~(16) (2013) 167401.
\newblock \href {https://doi.org/10.1103/PhysRevLett.110.167401}
  {\path{doi:10.1103/PhysRevLett.110.167401}}.

\bibitem{vasconcelos}
R.~Vasconcelos, S.~Reisenbauer, C.~Salter, G.~Wachter, D.~Wirtitsch,
  J.~Schmiedmayer, P.~Walther, M.~Trupke, {Scalable spin–photon entanglement
  by time-to-polarization conversion}, npj Quantum Inf. 6~(1) (2020) 9.
\newblock \href {https://doi.org/10.1038/s41534-019-0236-x}
  {\path{doi:10.1038/s41534-019-0236-x}}.

\bibitem{Chekhovich2020}
E.~A. Chekhovich, S.~F.~C. da~Silva, A.~Rastelli, {Nuclear spin quantum
  register in an optically active semiconductor quantum dot}, Nat. Nanotechnol.
  15~(12) (2020) 999--1004.
\newblock \href {https://doi.org/10.1038/s41565-020-0769-3}
  {\path{doi:10.1038/s41565-020-0769-3}}.

\bibitem{wang2012dynamicalcomparison}
Z.-H. Wang, G.~de~Lange, D.~Rist{\`{e}}, R.~Hanson, V.~V. Dobrovitski,
  {Comparison of dynamical decoupling protocols for a nitrogen-vacancy center
  in diamond}, Phys. Rev. B 85~(15) (2012) 155204.
\newblock \href {https://doi.org/10.1103/PhysRevB.85.155204}
  {\path{doi:10.1103/PhysRevB.85.155204}}.

\end{thebibliography}

\newpage
\appendix

\section{Cluster state equivalence}
\label{ap:state}

The following shows the evolution of the quantum state through one repetition ($N=1$) of the building block of the circuit of Fig.\,2, for a register of two nuclear spins coupled to the proxy qubit. We use the computational basis $\{\ket{0},\ket{1}\}$ for the spin qubits, and the circular polarisation basis $\{\ket{\text{L}},\ket{\text{R}}\}$ for the photon qubits (assuming a polarisation-based spin-photon entanglement scheme as example):

Initialise: \\
\begin{equation*}
\ket{1}\ket{1}\ket{1}
\end{equation*}

R$_y$($\pi$/2): \\
\begin{equation*}
(\ket{1}+\ket{0})\ket{1}\ket{1}
\end{equation*}

SWAP$_{12}$ \& R$_y$: \\
\begin{equation*}
(\ket{1}+\ket{0})(\ket{1}+\ket{0})\ket{1}
\end{equation*}

SWAP$_{12}$, SWAP$_{13}$ \& R$_y$: \\
\begin{equation*}
(\ket{1}+\ket{0})(\ket{1}+\ket{0})(\ket{1}+\ket{0})
\end{equation*}
 
SWAP$_{13}$: \& CZ$_{12}$: \\
\begin{equation*}
(-\ket{1}\ket{1}+\ket{1}\ket{0}+\ket{0}\ket{1}+\ket{0}\ket{0})(\ket{1}+\ket{0})
\end{equation*}

CZ$_{13}$: \\
\begin{equation*}
\begin{split}
&\ket{1}\ket{1}\ket{1}-\ket{1}\ket{0}\ket{1}+\ket{0}\ket{1}\ket{1}+\ket{0}\ket{0}\ket{1}+ \\ 
&-\ket{1}\ket{1}\ket{0}+\ket{1}\ket{0}\ket{0}+\ket{0}\ket{1}\ket{0}+\ket{0}\ket{0}\ket{0}
\end{split}
\end{equation*}

SWAP$_{12}$ \& photon generation: \\
\begin{equation*}
\begin{split}
&\ket{1}\ket{R}\ket{1}\ket{1}-\ket{0}\ket{L}\ket{1}\ket{1}+\ket{1}\ket{R}\ket{0}\ket{1}+ \\
&\ket{0}\ket{L}\ket{0}\ket{1}-\ket{1}\ket{R}\ket{1}\ket{0}+\ket{0}\ket{L}\ket{1}\ket{0}+ \\
&\ket{1}\ket{R}\ket{0}\ket{0}+\ket{0}\ket{L}\ket{0}\ket{0}
\end{split}
\end{equation*}

SWAP$_{12}$ \& photon generation: \\
\begin{equation*}
\begin{split}
&\ket{1}\ket{R}\ket{1}\ket{R}\ket{1}-\ket{1}\ket{R}\ket{0}\ket{L}\ket{1}+ \\
&\ket{0}\ket{L}\ket{1}\ket{R}\ket{1}+\ket{0}\ket{L}\ket{0}\ket{L}\ket{1}- \\
&\ket{1}\ket{R}\ket{1}\ket{R}\ket{0}+\ket{1}\ket{R}\ket{0}\ket{L}\ket{0}+ \\
&\ket{0}\ket{L}\ket{1}\ket{R}\ket{0}+\ket{0}\ket{L}\ket{0}\ket{L}\ket{0}
\end{split}
\end{equation*}

SWAP$_{13}$ \& photon generation: \\
\begin{equation*}
\begin{split}
&\ket{1}\ket{R}\ket{1}\ket{R}\ket{1}\ket{R}-\ket{1}\ket{R}\ket{0}\ket{L}\ket{1}\ket{R}+ \\
&\ket{1}\ket{R}\ket{1}\ket{R}\ket{0}\ket{L}+\ket{1}\ket{R}\ket{0}\ket{L}\ket{0}\ket{L}- \\
&\ket{0}\ket{L}\ket{1}\ket{R}\ket{1}\ket{R}+\ket{0}\ket{L}\ket{0}\ket{L}\ket{1}\ket{R}+ \\
&\ket{0}\ket{L}\ket{1}\ket{R}\ket{0}\ket{L}+\ket{0}\ket{L}\ket{0}\ket{L}\ket{0}\ket{L} 
\end{split}
\end{equation*}

SWAP$_{13}$ \& SWAP$_{12}$: \\
\begin{equation*}
\begin{split}
&\ket{1}\ket{R}\ket{1}\ket{R}\ket{1}\ket{R}-\ket{0}\ket{L}\ket{1}\ket{R}\ket{1}\ket{R}+ \\
&\ket{1}\ket{R}\ket{0}\ket{L}\ket{1}\ket{R}+\ket{0}\ket{L}\ket{0}\ket{L}\ket{1}\ket{R}- \\
&\ket{1}\ket{R}\ket{1}\ket{R}\ket{0}\ket{L}+\ket{0}\ket{L}\ket{1}\ket{R}\ket{0}\ket{L}+ \\
&\ket{1}\ket{R}\ket{0}\ket{L}\ket{0}\ket{L}+\ket{0}\ket{L}\ket{0}\ket{L}\ket{0}\ket{L} \\
\end{split}
\end{equation*}

R$_y$ \& SWAP$_{12}$: \\
\begin{equation*}
\begin{split}
&\ket{1}\ket{R}(\ket{1}+\ket{0})\ket{R}\ket{1}\ket{R}- \\
&\ket{1}\ket{R}(-\ket{1}+\ket{0})\ket{L}\ket{1}\ket{R}+ \\
&\ket{0}\ket{L}(\ket{1}+\ket{0})\ket{R}\ket{1}\ket{R}+ \\
&\ket{0}\ket{L}(-\ket{1}+\ket{0})\ket{L}\ket{1}\ket{R}- \\
&\ket{1}\ket{R}(\ket{1}+\ket{0})\ket{R}\ket{0}\ket{L}+ \\
&\ket{1}\ket{R}(-\ket{1}+\ket{0})\ket{L}\ket{0}\ket{L}+ \\
&\ket{0}\ket{L}(\ket{1}+\ket{0})\ket{R}\ket{0}\ket{L}+ \\
&\ket{0}\ket{L}(-\ket{1}+\ket{0})\ket{L}\ket{0}\ket{L}
\end{split}
\end{equation*}

R$_y$ \& SWAP$_{13}$: \\
\begin{equation*}
\begin{split}
&\ket{1}\ket{R}(\ket{1}+\ket{0})\ket{R}(\ket{1}+\ket{0})\ket{R}- \\
&\ket{1}\ket{R}(-\ket{1}+\ket{0})\ket{L}(\ket{1}+\ket{0})\ket{R}+ \\ 
&\ket{1}\ket{R}(\ket{1}+\ket{0})\ket{R}(-\ket{1}+\ket{0})\ket{L}+ \\
&\ket{1}\ket{R}(-\ket{1}+\ket{0})\ket{L}(-\ket{1}+\ket{0})\ket{L}- \\ 
&\ket{0}\ket{L}(\ket{1}+\ket{0})\ket{R}(\ket{1}+\ket{0})\ket{R}+ \\
&\ket{0}\ket{L}(-\ket{1}+\ket{0})\ket{L}(\ket{1}+\ket{0})\ket{R}+ \\
&\ket{0}\ket{L}(\ket{1}+\ket{0})\ket{R}(-\ket{1}+\ket{0})\ket{L}+ \\
&\ket{0}\ket{L}(-\ket{1}+\ket{0})\ket{L}(-\ket{1}+\ket{0})\ket{L}
\end{split}
\end{equation*}

R$_y$ \& SWAP$_{13}$: \\
\begin{equation*}
\begin{split}
&(\ket{1}+\ket{0})\ket{R}(\ket{1}+\ket{0})\ket{R}(\ket{1}+\ket{0})\ket{R}- \\
&(\ket{1}+\ket{0})\ket{R}(-\ket{1}+\ket{0})\ket{L}(\ket{1}+\ket{0})\ket{R}+ \\ 
&(-\ket{1}+\ket{0})\ket{L}(\ket{1}+\ket{0})\ket{R}(\ket{1}+\ket{0})\ket{R}+ \\
&(-\ket{1}+\ket{0})\ket{L}(-\ket{1}+\ket{0})\ket{L}(\ket{1}+\ket{0})\ket{R}- \\
&(\ket{1}+\ket{0})\ket{R}(\ket{1}+\ket{0})\ket{R}(-\ket{1}+\ket{0})\ket{L}+ \\
&(\ket{1}+\ket{0})\ket{R}(-\ket{1}+\ket{0})\ket{L}(-\ket{1}+\ket{0})\ket{L}+ \\
&(-\ket{1}+\ket{0})\ket{L}(\ket{1}+\ket{0})\ket{R}(\ket{1}-\ket{0}\ket{L})+ \\
&(-\ket{1}+\ket{0})\ket{L}(-\ket{1}+\ket{0})\ket{L}(-\ket{1}+\ket{0})\ket{L}
\end{split}
\end{equation*}

Measure, assuming a collapse of all spin qubits into the $\ket{1}$ state: \\
\begin{equation*}
\begin{split}
&\ket{R}\ket{R}\ket{R}-\ket{R}\ket{L}\ket{R}+\ket{L}\ket{R}\ket{R}+\ket{L}\ket{L}\ket{R}- \\
&\ket{R}\ket{R}\ket{L}+\ket{R}\ket{L}\ket{L}+\ket{L}\ket{R}\ket{L})+\ket{L}\ket{L}\ket{L}
\end{split}
\end{equation*}

As desired this is a three-photon cluster state.

\section{Group IV Nuclear control}
\label{ap:g4hamil}

We consider the $S=\frac{1}{2}$ electronic state of a negative group IV vacancy which is coupled to its intrinsic nuclear spin via a hyperfine interaction. As mentioned in the main text, this coupling has been observed to be mostly isotropic and thus the Hamiltonian for this system can be written as,

\begin{align*}
H = \gamma_e B_z  \frac{\sigma_z}{2} \otimes I + \frac{\Omega}{2} \cos(\omega t) \sigma_x \otimes I + \gamma_n I \otimes \frac{\vec{B} \cdot \vec{\sigma}}{2}\\
+ A_\parallel \frac{\sigma_z}{2} \otimes \frac{\sigma_z}{2} + A_\perp (\frac{\sigma_x}{2} \otimes \frac{\sigma_x}{2} + \frac{\sigma_y}{2} \otimes \frac{\sigma_y}{2})\text{,}    
\end{align*}

\noindent
where $\vec{B}$ is the magnetic field in a coordinate frame where $\hat{z}$ is oriented along the symmetry axis of the defect and $\gamma_e$ and $\gamma_n$ denote the electronic and nuclear gyromagnetic ratios respectively. Here we have assumed that $\gamma_e B_\text{x,y} \ll \lambda_\text{SO}$ where $\lambda_\text{SO}$ is the spin-orbit coupling and thus the electronic quantization axis is approximately $\hat{z}$. For SiV, its ground state splitting of 50GHz corresponds to an effective magnetic field on the electron of 2T; for SnV, it is 30T. $A_\parallel$ and $A_\perp$ are the parallel and perpendicular components of the hyperfine interaction and $\Omega$ is the Rabi frequency, here taken as real, and $\omega$ the microwave frequency. 
To obtain a time-independent problem we introduce a rotating frame at frequency $\omega$ by performing the transformation $e^{-i \omega t \sigma_z/2 \otimes I} H e^{i \omega t \sigma_z/2 \otimes I}$ and define the detuning $\delta = \omega - \gamma_e B_z$. Neglecting the non-secular $A_\perp$ term the new Hamiltonian,

\begin{equation}
H = \delta \frac{\sigma_z}{2} \otimes I + \Omega \frac{\sigma_x}{2} \otimes I + A_\parallel \frac{\sigma_z}{2} \otimes \frac{\sigma_z}{2} + \gamma_n I \otimes \frac{\vec{B} \cdot \vec{\sigma}}{2}\text{,}
\label{eq:hamiltonian}
\end{equation}

\noindent
can be exploited to control the nuclear spin via the electron and thus to perform two-qubit gates. We propose to work in a regime where the magnetic field is off-axis and therefore the nuclear precession axis is dependent on the electronic state, due to the hyperfine interaction. While this off-axis magnetic field enables conditional control of the nuclear spin, it also leads to unwanted spin-flipping transitions when decaying from the excited state. A high degree of cyclicity is nevertheless preserved when spin-orbit dominates, as is the case for heavier Group IV elements or weak off-axis magnetic fields.

\section{Dynamical decoupling gates}
\label{ap:simulation}

An off-axis magnetic field introduces terms in the Hamiltonian of the form $\sigma_z \otimes \sigma_x$. These are analogous to the dipolar coupling terms between electronic spin qubits and neighbouring nuclei, for which control has been demonstrated experimentally in NVs and more recently in Group IV defects \cite{taminiau2014universalcontrol, bhaskar2020memoryenhanced}. Following equation \ref{eq:hamiltonian} for the case of an off-axis magnetic field component along $\hat{x}$, the nuclear spin will precess along $\vec{\omega_{\pm}} = \left( \gamma_n B_x / 2, 0, \gamma_n B_z / 2 \pm \frac{A_\parallel}{4} \right)$ for the electron in the states $S = \pm 1/2$. This conditional precession can be exploited via dynamical decoupling sequences of the form $(\tau_f - \pi - \tau_f)^N$ by tuning the interpulse spacing 2$\tau_f$ in resonance with the nucleus dynamics. 

The nature of the resonances depends on the relative values of $\gamma_n B_x$ and $A_\parallel$. For the case of weakly coupled $^{13}\text{C}$, $\gamma_n B_x/2 \gg A_\parallel$ and the precession axes are close to parallel ($\hat{\omega}_{+} \cdot \hat{\omega}_{-} \sim 1$). For strongly coupled nuclei, like $^{29}\text{Si}$, $A_\parallel$ typically dominates and thus the axis are close to anti-parallel ($\hat{\omega}_{+} \cdot \hat{\omega}_{-} \sim -1$). In this case, conditional rotations of the nuclear qubit are obtained when

\begin{equation*}
(|\vec{\omega_+}| - |\vec{\omega_-}|) \tau = (2n-1) \pi \text{,}
\label{eq:resonance}
\end{equation*}
where $n$ is a positive integer. Unconditional rotations along two perpendicular axis are obtained under a similar resonance condition where $2n-1$ is replaced with $2n$ and when the system is driven off-resonantly. This set of gates, along with control of the electron, is universal and can be utilised to construct the required SWAP and CZ gates. 

In our simulation, we construct the SWAP and CZ gates by numerically finding the combination of $\tau_f - \pi - 2\tau_f - \pi - \tau_f$ units that results in the highest fidelity. To do so, we select a target gate and start by concatenating $k$ of these units with interpulse spacings $\{\tau_{f,0}, \tau_{f,1}, ..., \tau_{f,k}\}$. To allow for the generation of any two-qubit gate, we also include, at random, electron gates between the dynamically decoupled units. By perturbing these spacings and computing the overall unitary transformation of the gate for an ideal system with no decoherence, we find the local maximum of the gate fidelity with respect to all $\tau_i$. This local optimisation is run repeatedly for variable $k$ until a gate is found that meets the required fidelity threshold of $99.9\%$.

To model the effects of decoherence, an additional term is added to the Hamiltonian, of the form:

$$H_\text{bath} = B(t) \frac{\sigma_z}{2} \otimes I \text{.}$$

\noindent
The noise field $B(t)$ can take several forms. The simplest assumption is that of a stationary Gaussian and Markovian noise source: an Ornstein–Uhlenbeck process. This can be physically motivated through a bath of nearby nuclear spins. Analytical relations can be derived relating the strength ($b$) and timescale of correlations ($\tau_c$) to measurements of the inhomogeneous dephasing $T_2^*$ and the coherence time $T_2$ of a Hahn-echo \cite{wang2012dynamicalcomparison}

$$T_2^* = \frac{\sqrt{2}}{b}, \quad  T_2 = \left( \frac{12 \tau_c}{b^2} \right) ^{1/3} \text{.} $$

\end{document}